\documentclass{aa}
\usepackage{psfig}
\usepackage{natbib}
\usepackage{graphics}

\bibpunct{(}{)}{;}{a}{}{,}
\defcitealias{ity97}{ITY97}
\defcitealias{han98}{HAN98}
\defcitealias{dsb+98}{DSBH98}

\newcommand{\msun}{\mbox{${\rm M}_{\sun}$}}

\begin{document}

   \thesaurus{06            
              (08.02.1;     
               08.05.3;     
               08.19.1;     
               08.23.1)}    
             
 \title{Population synthesis for double white dwarfs\\ I.
               Close detached systems}

\author{Gijs Nelemans\inst{1}, Lev R. Yungelson\inst{1,2}, Simon F.
  Portegies Zwart\inst{3}\thanks{Hubble Fellow} and Frank
  Verbunt\inst{4}} \offprints{G. Nelemans, gijsn@astro.uva.nl}

\institute{
     Astronomical Institute ``Anton Pannekoek'', 
     Kruislaan 403, NL-1098 SJ Amsterdam, the Netherlands; gijsn@astro.uva.nl 
             \and
     Institute of Astronomy of the Russian Academy of
     Sciences, 48 Pyatnitskaya Str., 109017 Moscow, Russia; lry@inasan.rssi.ru 
             \and
     Department of Physics and Center for Space Research, MIT,
     77 Massachusetts Avenue, Cambridge MA 02139 USA; spz@space.mit.edu
                \and 
     Astronomical Institute, Utrecht University,
     P.O.Box 80000, NL-3508 TA Utrecht, the Netherlands; F.W.M.Verbunt@astro.uu.nl
}

\date{received July 3, 2000}
\maketitle

\markboth{G. Nelemans et al. : Close detached double white dwarfs}{}

\begin{abstract}
  
  We model the population of double white dwarfs in the Galaxy and
  find a better agreement with observations compared to earlier
  studies, due to two modifications. The first is the treatment of the
  first phase of unstable mass transfer and the second the modelling
  of the cooling of the white dwarfs.
  
  A satisfactory agreement with observations of the local sample of
  white dwarfs is achieved if we assume that the initial binary
  fraction is $\sim$50 \% and that the lowest mass white dwarfs ($M <
  0.3 \msun$) cool faster than the most recently published cooling
  models predict.

  With this model we find a Galactic birth rate of close double
  white dwarfs of 0.05 yr$^{-1}$, a birth rate of AM CVn systems of
  0.005 yr$^{-1}$, a merger rate of pairs with a combined mass
  exceeding the Chandrasekhar limit (which may be progenitors of
  SNe~Ia) of 0.003 yr$^{-1}$ and a formation rate of planetary nebulae
  of 1 yr$^{-1}$.  We estimate the total number of double white dwarfs
  in the Galaxy as 2.5\,$\times\,10^8$.
  
  In an observable sample with a limiting magnitude $V_{\rm lim} = 15$
  we predict the presence of $\sim$855 white dwarfs of which
  $\sim$220 are close pairs. Of these 10 are double CO white
  dwarfs of which one has a combined mass exceeding the Chandrasekhar
  limit and will merge within a Hubble time.
  
\end{abstract}

\section{Introduction}

Close double white dwarfs\footnote{Throughout this work we'll
  use the term double white dwarf instead of double degenerate, which
  is commonly used, because the term double degenerate is sometimes
  used for white dwarf - neutron star or double neutron star
  binaries.}  form an interesting population for a number of reasons.
First they are binaries that have experienced at least two phases of
mass transfer and thus provide good tests for theories of binary
evolution.  Second it has been argued that type Ia supernovae arise
from merging double CO white dwarfs \citep{web84,it84a}.  Thirdly
close double white dwarfs may be the most important contributors to
the gravitational wave signal at low frequencies, probably even
producing an unresolved noise burying many underlying signals
\citep{eis87,hbw90}.  A fourth reason to study the population of
double white dwarfs is that in combination with binary evolution
theories, the recently developed detailed cooling models for
(low-mass) white dwarfs can be tested.

The formation of the population of double white dwarfs has been
studied analytically by \citet{it86a,it87} and numerically by
\citet{lp88,ty93,ty94,ylt+94,hpe95}; \citet[hereafter
\citetalias{ity97}]{ity97}, and \citet[hereafter
\citetalias{han98}]{han98}.  Comparison between these sudies gives
insight in the differences that exist between the assumptions made in
different synthesis calculations.

Following the discovery of the first close double white dwarf
\citep{slo88}, the observed sample of such systems in which the
  mass of at least one component is measured has increased to 14
\citep{mm99,mmm+00}. This makes it possible to compare the models to
the observations in more detail.

In this paper we present a new population synthesis for double white
dwarfs, which is different from previous studies in three aspects.
The first are some differences in the modelling of the binary
evolution, in particular the description of a common envelope without
spiral-in, in which the change in orbit is governed by conservation of
angular momentum, rather than of energy (Sect.~\ref{binev}). The
second new aspect is the use of detailed models for the cooling of
white dwarfs (Sect.~\ref{cooling}), which are important because it is
the rate of cooling which to a large extent determines how long a
white dwarf remains detectable in a magnitude-limited observed sample.
The third new aspect is that we use different models of the star
formation history (Sect.~\ref{SFH}).  Results are presented in
Sect.~\ref{results} and discussed in Sect.  \ref{discussion}. The
conclusions are summarised in Sect.~\ref{conclusion}. In the Appendix
some details of our population synthesis are described .

\section{Binary and single star evolution; the formation of double
  white dwarfs}\label{binev}

The code we use is based on the code described by \citet{pv96} and
\citet{py98}, but has been modified in two respects; the white dwarf
masses and the treatment of unstable mass transfer.

\subsection{White dwarf masses}\label{wdmass}

The masses of white dwarfs in binaries provide important observational
constraints on evolution models. Therefore we have improved the
treatment of the formation of white dwarfs in our binary evolution
models by keeping more accurate track of the growth of the mass of the
core. Details are given in Appendix~\ref{a:wdmass}.

\subsection{Unstable mass transfer}\label{masstransfer}

There exist two ``standard'' scenarios for the formation of close
double white dwarfs. In the first, the binary experiences
two stages of unstable mass transfer in which a common envelope is
formed. The change of the binary orbital separation in a common
envelope is treated on the base of a balance between orbital energy
and the binding energy of the envelope of the mass-losing star
\citep{pac76,ty79,web84,il93}. The second scenario assumes that the
first-born white dwarf of the pair is formed via stable mass transfer,
like in Algol-type binaries (possibly accompanied by some loss of mass
and angular momentum from the system) and the second white dwarf is
formed via a common envelope.

Reconstruction of the evolution of three double helium white dwarfs
with known masses of both components led us to the conclusion that a
spiral-in could be avoided in the first phase of unstable mass
transfer \citep{nvy+00}.  Briefly, when the mass ratio of two stars
entering a common envelope is not too far from unity, we assume that
the envelope of the evolving giant is ejected without a spiral-in, and
that the change in orbital separation is governed by conservation of
angular momentum (the equation used is given in
Appendix~\ref{a:dynamic}).  We parametrise the loss of angular
momentum from the binary with a factor $\gamma$.  If the mass ratio is
more extreme, the common envelope leads to a spiral-in, which is
governed by the conservation of energy (the equation used is given in
Appendix~\ref{a:stace}). The efficiency with which the energy of the
binary orbit is used to expell the envelope of the giant is
parametrised by a factor $\alpha_{\rm ce}\lambda$.  We switch between
the two descriptions at the mass ratio where both give the same change
of the separation (roughly at 0.2).  \citet{nvy+00} find that values
of $\gamma = 1.75$ and $\alpha_{\rm ce} \lambda = 2$ give the best
agreement of evolution models with the observed parameters of three
binaries in which the masses of both white dwarfs are known, and
therefore we use these values in our calculations.

Another novelty is what we suggest to call {\em ``double spiral-in''}.
It describes the situation when the primary fills its Roche lobe at
the time that its companion has also evolved off the main sequence.
This kind of evolution can only take place when the initial mass ratio
is close to unity. Such a mass transfer phase has hitherto been
described with the standard common envelope formalism; in the same way
as when the companion is still a main sequence star.  However, if the
companion is evolved, one might as well argue that the envelope of the
smaller star becomes part of the common envelope, and the envelopes of
{\em both} stars will be expelled.  We propose to use the energy
balance here, since the double core binary will in general not have
enough angular momentum to force the envelope into co-rotation.
An equation for the change in orbital separation in the case of a
``double spiral-in'' is derived in Appendix~\ref{a:double_spi} exactly
analogous to the usual common envelope formalism
\citep[e.g.][]{web84}.

\subsection{Examples}\label{examples}

\begin{figure*}[t]
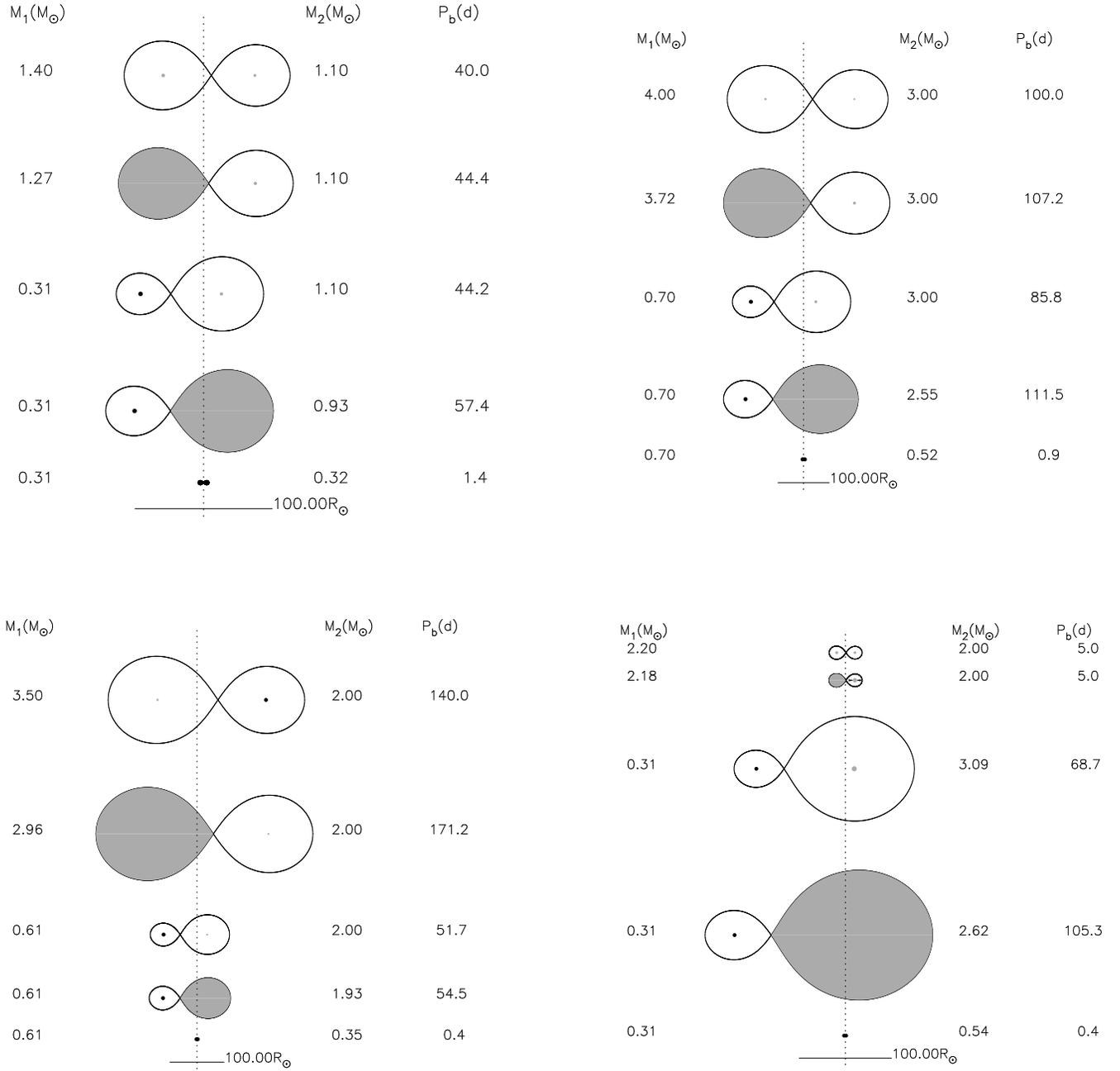

\begin{minipage}{0.45\textwidth}
\centerline{\psfig{figure=H2305f1.ps,angle=-90,width=\columnwidth,bbllx=550pt,bblly=670pt,bburx=20pt,bbury=240pt,clip=}}
\end{minipage}
\hspace*{0.095\textwidth}
\begin{minipage}{0.45\textwidth}
\centerline{\psfig{figure=H2305f2.ps,angle=-90,width=\columnwidth,bbllx=550pt,bblly=710pt,bburx=20pt,bbury=230pt,clip=}}
\end{minipage}
\\
\begin{minipage}{0.45\textwidth}
\centerline{\psfig{figure=H2305f3.ps,angle=-90,width=\columnwidth,bbllx=550pt,bblly=700pt,bburx=20pt,bbury=210pt,clip=}}
\end{minipage}
\hspace*{0.095\textwidth}
\begin{minipage}{0.45\textwidth}
\centerline{\psfig{figure=H2305f4.ps,angle=-90,width=\columnwidth,bbllx=550pt,bblly=710pt,bburx=20pt,bbury=210pt,clip=}}
\end{minipage}
\caption[]{Evolutionary scenarios for the formation of a double helium
  white dwarf (top left), a double CO white dwarf (top right) and the
  CO+He and He+CO pairs (bottom ones). Note that the scales in the
  panels differ as indicated by the 100 R$_{\sun}$ rulers at the
  bottom. For a more detailed discussion see Sect.~\ref{examples}}
\label{fig:examples}
\end{figure*}

Before discussing effects that influence the double white dwarf
population as a whole we discuss some typical examples of binary
evolution leading to close double white dwarfs, to illustrate some of
the assumptions used in our models.  For details of the treatment of
binary evolution we refer to \citet{pv96} and the Appendix.

\subsubsection{Double helium white dwarfs}

The most common double white dwarfs consist of two helium white dwarfs
(Sect.~\ref{birth-rates-numbers}). These white dwarfs descend from
systems in which both stars have $M \la$ 2.3\,\msun\ and fill their
Roche lobes before He ignition in their degenerate cores.  In
Fig.~\ref{fig:examples} (top left) we show an example of the formation
of such a system.  We start with a binary with an orbital period of 40
days and components of 1.4 and 1.1\,\msun.  The primary fills its
Roche lobe after 3 Gyrs, at which moment it has already evolved up the
first giant branch and has lost $\sim$0.13\,\msun\ in a stellar wind.
When the star fills its Roche lobe it has a deep convective envelope,
so the mass transfer is unstable. We apply the envelope ejection
formalism to describe the mass transfer with a $\gamma$-value of 1.75
(see Eq.~\ref{eq:orbit_change}).  The core of the donor becomes a
0.31\,\msun\ helium white dwarf.  The orbital period of the system
hardly changes. After 4 Gyr, when the first formed white dwarf has
already cooled to very low luminosity, the secondary fills its Roche
lobe and has a deep convective envelope. Mass loss again proceeds on
dynamical time scale, but the mass ratio of the components is rather
extreme and a common envelope is formed in which the orbit shrinks
dramatically.

\subsubsection{Double CO white dwarfs}

Most double CO white dwarfs are formed in systems which are initially
so wide that both mass transfer phases take place when the star is on
the AGB and its core consists already of CO, such that a CO white
dwarfs are formed directly.  An example is shown in
Fig.~\ref{fig:examples} (top right). In the first phase of mass
transfer the change of the orbital separation is regulated by the
conservation of angular momentum during envelope ejection, according
to equation (Eq.~\ref{eq:orbit_change}), while in the second phase of
mass transfer spiral-in is described by Eq.~(\ref{eq:ce}).

Much less frequently, CO white dwarfs are formed by stars more massive
than 2.3\,\msun\ which fill their Roche lobe when they have a
nondegenerate core, before helium ignition.  Roche lobe overflow then
results in the formation of a low-mass helium star.  A brief
additional phase of mass transfer may happen, if the helium star
expands to giant dimensions during helium shell burning. This is the
case for 0.8\,$\la M_{\rm He}/\msun \la$\,3 (see
Appendix~\ref{a:helium_stars}). After exhaustion of helium in its
core, the helium star becomes a CO white dwarf.

\subsubsection{CO white dwarfs with He companions}\label{examples:CO+He}

In Fig.~\ref{fig:examples} (bottom left) we show an example in which
the CO white dwarf is formed first. It starts with a more extreme mass
ratio and a relatively wide orbit, which shrinks in a phase of
envelope ejection. The secondary does not accrete anything and fills
its Roche lobe when it ascends the first giant branch, having a
degenerate helium core. It then evolves into a helium white dwarf.

In the second example (shown in Fig.~\ref{fig:examples}; bottom
right), the system evolves through a stable mass exchange phase
because the primary has a radiative envelope when it fills its Roche
lobe. Part of the transferred mass is lost from the systems (see
  Appendix~\ref{a:stable}). The orbit widens and the primary forms a
helium white dwarf when it has transferred all its envelope to its
companion.  The secondary accretes so much mass that it becomes too
massive to form a helium white dwarf.  The secondary fills its Roche
lobe on the AGB to form a CO white dwarf in a common envelope in which
the orbital separation reduces strongly. Because of the differential
cooling (Sec.~\ref{cooling}) the CO white dwarf, despite the fact that
it is formed last, can become fainter than its helium companion.
Since the probability to fill their Roche lobe when the star has a
radiative envelope, is low for low-mass stars, the scenario in which
the helium white dwarf is formed first is less likely (see
Sect.~\ref{results}).

\section{A model for the current population of white dwarfs in the
Galaxy}

\begin{table}[t]
\caption[]{Models and their parameters. IMF is always according to
  \citet{ms79}. The SFR is either exponentially
  decaying (Eq.~\ref{eq:sfr}) or constant. The column ``\% binaries'' gives the
  initial binary fraction in the population, the column ``cooling'' gives 
  the  cooling model (see Sect.~\ref{cooling})}
\label{tab:models}
\begin{center}
\begin{tabular}{lrrrr}
Model & SFH    & \% binaries & cooling & \\ \hline
A1    & Exp    & 50          & DSBH98 & \\           
A2    & Exp    & 50          & Modified DSBH98  &  \\  
A3    & Exp    & 50          & 100 Myr &  \\ 
B     & Exp    & 100         & Modified DSBH98 &  \\ 
C     & Cnst   & 50          & Modified DSBH98 &  \\ 
D     & Cnst   & 100         & Modified DSBH98 & \\ \hline
\end{tabular}
\end{center}
\end{table}

We model the current population of double and single white dwarfs in
the Galaxy using population synthesis and compare our models with the
observed populations. We initialise 250,000 ``zero-age'' binaries and
evolve these binaries according to simplified prescriptions for single
and binary star evolution, including stellar wind, mass transfer
(which may involve loss of mass and angular momentum from the binary),
common envelopes and supernovae.

For each initial binary the mass $M_{\rm i}$ of the more massive
component, the mass ratio $q_{\rm i} \equiv m_{\rm i}/M_{\rm i} \le
1$, where $m_{\rm i}$ is the mass of the less massive component, the
orbital separation $a_{\rm i}$ and eccentricity $e_{\rm i}$ are chosen
randomly from distributions given by
\begin{eqnarray}
\label{eq:init}
{\rm  Prob}(M_{\rm i}) \quad & \mbox{MS79}  
&  {\rm for} \; \; \; 0.96 \, \msun \leq M_{\rm i} \leq 11\,\msun, \nonumber\\ 
{\rm Prob}(q_{\rm i}) \propto & {\rm const.} & {\rm for} \; \; \; \; 0
< q_{\rm i} \leq 1,  \nonumber\\
{\rm Prob}(a_{\rm i}) \propto & a_{\rm i}^{-1} & {\rm for} \; \; \; \; 0 \leq
\log a_{\rm i}/R_{\sun} \leq 6, \\ 
{\rm Prob}(e_{\rm i}) \propto & 2 e_{\rm i} & {\rm for} \; \; \; \; 0
\leq e_{\rm i} \leq 1.\nonumber 
\end{eqnarray}
For the primary mass we use the approximation of \citet{eft89} to the
\citet{ms79} IMF indicated as MS79. A primary at the lower mass limit
has a main sequence life time equal to our choice of the age of the
Galactic disk (10 Gyr).  The lower mass of less massive component is
set to 0.08\,\msun, the minimum mass for hydrogen core burning.  The
distribution over separation is truncated at the lower end by the
separation at which the ZAMS binary would be semi-detached.

To investigate the effects of different cooling models
(Sect.~\ref{cooling}) and different assumptions about the star
formation history (Sect.~\ref{SFH}) different models have been
computed (Table~\ref{tab:models}).

\section{Modelling the observable population; white dwarf cooling}\label{cooling+selection}

To model the observable population we have to take orbital evolution 
and selection effects into account.

\subsection{Orbital evolution of double white dwarfs}\label{evolution}

The most important effect of orbital evolution, which is taken into
account also in all previous studies of close binary white dwarfs, is
the disappearance from the sample of the tightest systems as they
merge, due to the loss of angular momentum via gravitational wave
radiation.  For example an 0.6\,\msun + 0.6\,\msun\ white dwarf pair
with orbital period of 1 hour merges in $3\,\times\,10^7$\,yr. If it
is located at a distance of 100\,pc from the Sun it will disappear
abruptly from a magnitude limited sample by merging \footnote{Note,
  however, that just before merging white dwarfs may become quite
  bright due to tidal heating \citep{itf98}.} before the white dwarfs
have become undetectable due to cooling.

\subsection{Selection effects}\label{selection}

The observed double white dwarfs are a biased sample. First, they were
mainly selected for study because of their supposed low mass, since
this is a clear indication of binarity \citep{slo88,mdd95}. Secondly,
for the mass determinations and the measurement of the radial
velocities the white dwarfs must be sufficiently bright. A third
requirement is that the radial velocities must be large enough that
they can be found, but small enough that spectral lines don't get smeared
out during the integration. \citet{mm99} discuss
this last requirement in detail.  Following them, we include a detection
probability in the model assuming that double white dwarfs in the
orbital period range between 0.15\,hr and 8.5\,day will be detected
with 100\% probability and that above 8.5\,day the detection
probability decreases linearly from 1 at 8.5 days to 0 at $\sim$35
days \citep[see Fig. 1 in][]{mm99}.

The second selection effect is related to the brightness of the white
dwarfs, which is governed by their cooling curves. 

\subsection{White dwarf cooling}\label{cooling}

\citet{it85} noticed that for a 0.6\,\msun\ white dwarf the maximum
probability of discovery corresponds to a cooling age of $\sim
10^8$\,yr. In absence of detailed cooling curves for low-mass white
dwarfs, it was hitherto assumed in population synthesis studies that
white dwarfs remain bright enough to be observed during $10^8$\,yr,
irrespective of their mass. However, recent computations (Bl\"ocker
\citeyear{blo95b}; Driebe et al. \citeyear{dsb+98}, hereafter
\citetalias{dsb+98}; Hansen \citeyear{han99}) indicate that helium
white dwarfs cool more slowly than CO white dwarfs, for two reasons.
First, helium cores contain a higher number of ions than carbon-oxygen
cores of the same mass, they store more heat and are brighter at the
same age \citep{han99}. Second, if the mass of the hydrogen envelope
of the white dwarf exceeds a critical value, pp-reactions remain the
main source of energy down to effective temperatures well below $10^4$
K (Webbink \citeyear{web75}; \citetalias{dsb+98}; Sarna~et~al.
\citeyear{sea00}).  This residual burning may lead to a significant
slow-down of the cooling.

White dwarfs in close binaries form when the evolution of (sub)giants
with degenerate cores and hydrogen-rich envelopes is terminated by
Roche lobe overflow. The amount of hydrogen that is left on the white
dwarf depends on the details of this process. Fully fledged
evolutionary calculations of the formation of helium white dwarfs,
e.g. \citet{gg70,sea00}, as well as calculations that mimic Roche
lobe overflow by mass loss at fixed constant rate \citep{dsb+98}, find
that the thickness of the residual envelope around the white dwarf is
increasing with decreasing white dwarf mass. As a result the
brightness at fixed age decreases monotonically with increasing white
dwarf mass (see also Fig.~\ref{fig:wdcooling}). 

However, it is not clear that these calculations are valid for white
dwarfs formed in a common envelope.  In addition, white dwarfs may
lose mass by stellar wind when they still have a high luminosity. Such
winds are observed for nuclei of planetary nebulae and post-novae and
could also be expected for He white dwarfs.  Finally, white dwarfs
with masses between $\sim$0.2 and $\sim$0.3\,\msun\ experience thermal
flashes \citep{ktw68,web75,it86b,dbs+99,sea00}, in which the envelopes
expand. This may lead to additional mass loss in a temporary common
envelope, especially in the closest systems with separations $\la
1\,R_{\sun}$.  Mass loss may result in extinguishing of hydrogen burning
\citep{it86b,sea00}. 

\citet{han99} argues that the details of the loss of the hydrogen
envelope are very uncertain and assumes that all white dwarfs have a
hydrogen envelope of the same mass. He finds that helium white dwarfs
cool slower than the CO white dwarfs, but inside these groups, the
more massive white dwarfs cool the slowest. The difference within the
groups are small.

We conclude that the cooling models are still quite uncertain, so we
will investigate the result of assuming different cooling models in
our population synthesis.

The first model we compute (A1; see Table~\ref{tab:models} for a list
of all computed models) uses the cooling curves as given by
\citet{blo95b} for CO white dwarfs and \citetalias{dsb+98} for He
white dwarfs as detailed in Appendix \ref{a:wd_cooling}.  For the
second model (A2) we made a crude estimate of the cooling curves for
the case that the thermal flashes or a stellar wind reduce the mass of
the hydrogen envelope and terminate the residual burning of hydrogen.
We apply this to white dwarfs with masses below 0.3\,\msun, and model
all these white dwarfs identically and simply with cooling curves for
a more massive (faster cooling) white dwarf of 0.46 \msun.
To compare with the previous investigators, we include one model (A3)
in which all white dwarfs can be seen for 100 Myrs. We did not model
the cooling curves of \citet{han99}, because no data for $L > 0.01
L_{\sun}$ are given.

\subsection{Magnitude limited samples and local space densities}\label{rho}

To convert the total Galactic population to a local population and to
compute a magnitude limited sample, we assume a distribution of all
single and binary stars in the galactic disk of the form
\begin{equation}\label{eq:rho}
\rho(R, z) = \rho_{\rm 0} \; e^{-R/H} \; \mbox{sech}(z/h)^2  \quad \mbox{pc}^{-3}
\end{equation}
where we use $H$ = 2.5 kpc \citep{sac97} and $h$ = 200 pc,
neglecting the age and mass dependence of $h$.

To construct a magnitude limited sample, we compute the magnitude for
all model systems from the cooling curves and estimate the
contribution of each model system from Eq.~(\ref{eq:rho}). The
absolute visual magnitudes along the cooling curves are derived using
bolometric corrections after \citet{eft89}.

From Eq.~(\ref{eq:rho}) the local ($R = 8.5 \mbox{kpc}, z = 30
\mbox{pc}$) space density ($\rho_{i,\sun}$) of any type of system is
related to the total number in the Galaxy ($N_i$) by:
\begin{equation}\label{eq:rho_sun}
\rho_{i,\sun} = N_i/ 4.8 \, \times \,10^{11} \quad \mbox{pc}^{-3}.
\end{equation}

\section{Star formation history}\label{SFH}

Some progenitors of white dwarfs are formed long ago. Therefore the
history of star formation in the Galaxy affects the contribution of
old stars to the population of local white dwarfs. To study this we
compute different models.

For models A and B (see Table~\ref{tab:models}), we model the star
formation history of the galactic disk as
\begin{equation}\label{eq:sfr}
{\rm SFR}(t) = 15 \; \exp(-t/\tau) \quad \msun \; \mbox{yr}^{-1}
\end{equation}
where $\tau$ = 7 Gyr. It gives a current rate of 3.6\,\msun\,yr$^{-1}$
which is compatible with observational estimates \citep{ran91,hj97}.
The integrated SFR, i.e.  the amount of matter that has been turned
into stars over the whole history of the galactic disk (10 Gyr) with
this equation is $\sim$8\,$\times\,10^{10}$ \msun\, which is higher
than the current mass of the disk, since part of the gas that is
turned into stars is given back to the ISM by supernovae and stellar
winds.

For models C and D we use a constant SFR of $4\,\msun$\ yr$^{-1}$
\citep[as][]{ty93}. We use an age of the disk of 10 Gyr, while
\citet{ty93} use 15 Gyr. Model D also allows us to compare our results
with previous studies (\citetalias{ity97} and \citetalias{han98}; see
Sect.~\ref{discussion}).

Most binary population synthesis calculations take a binary fraction
of 100\%. Since we want to compare our models with the observed
fraction of close double white dwarfs among all white dwarfs, we
present models with 100\% binaries (models B and D); and with 50\%
binaries and 50\% single stars, i.e. with 2/3 of all stars in binaries
(models A and C).

\section{Results}\label{results}

\begin{table*}[t]
\begin{minipage}{0.6\textwidth}
\begin{tabular}{lrrrrrrr}
  Model & SFH & \% bin & $\nu_{\rm (wd, wd)}$&$\nu_{\rm merge}$&
  SN~Ia&$\nu_{\rm AM CVn}$&\#(wd, wd)\\ 
  & & &($10^{-2}$)&($10^{-2}$)&($10^{-3}$)&($10^{-3}$)&($10^8$)\\ \hline
  A       & Exp & 50  & 4.8 & 2.2~~ & 3.2 & 4.6~ & 2.5~~ \\  
  B       &Exp  & 100 & 8.1 & 3.6~~ & 5.4 & 7.8~ & 4.1~~ \\ 
  C       &Cnst& 50  & 3.2 & 1.6~~ & 3.4 & 3.1~ & 1.2~~  \\
  D       &Cnst& 100 & 5.3 & 2.8~~ & 5.8 & 5.2~ & 1.9~~ \\ 
  \citetalias{ity97}$^1$&Cnst &100& 8.7 & 2.4~~ & 2.7 & 12.0 & 3.5~~ \\
  \citetalias{han98}$^1$&Cnst& 100& 3.2 & 3.1~~ & 2.9 & 26~~~~& 1.0~~ \\ \hline
\end{tabular}\\
$^1$ Note that \citetalias{ity97} and \citetalias{han98} used a
normalisation that is higher than we use for model D by factors
$\sim$1.4 and $\sim$1.1 respectively (see
Sect.~\ref{discussion:birthrates}).\\
\end{minipage}
\hspace*{0.03\textwidth}
\begin{minipage}{0.32\textwidth}
\caption[]{Birth and event rates and numbers for the different models. All
  birth and event rates ($\nu$) are in units of yr$^{-1}$ in the
  Galaxy. All numbers (\#) are total numbers in the Galaxy. Close
  double white dwarfs are represented with (wd, wd). See
  Sect.~\ref{birth-rates-numbers} for a discussion of these rates. For
  comparison (in Sect.~\ref{discussion:birthrates}) we also include
  numbers computed by the code from \citetalias{ity97} but using an
  age of the galactic disk of 10 Gyr instead of the 15 Gyr used by
  \citetalias{ity97}; and numbers of model 1 of \citetalias{han98}.}
\label{tab:birthrates}
\end{minipage}
\end{table*}

Our results are presented in the next five subsections. In
Sect.~\ref{birth-rates-numbers} we give the birth rates and total
number of double white dwarfs in the Galaxy. These numbers allow a
detailed comparison with results of earlier studies, which we defer to
Sect.~\ref{discussion}. They cannot be compared with observations
directly, with the exception of the SN~Ia rate. For comparison with
the observed sample, described in Sect.~\ref{obssample}, we compute
magnitude limited samples in the remaining sections.  In
Sect.~\ref{results:P_m} the distribution over periods and masses is
compared with the observations, which constrains the cooling models.
Comparison of the mass ratio distribution with the observations gives
further support for our new description of a common envelope without
spiral-in (Sect.~\ref{sec:p-q}). In Sect.~\ref{wd_mass} we compare our
model with the total population of single and binary white dwarfs and
in Sect.~\ref{PN} we compare models that differ in the assumed star
formation history with the observed rate of PN formation and the local
space density of white dwarfs.

\subsection{Birth rates and numbers}\label{birth-rates-numbers}

In Table~\ref{tab:birthrates} the birth rates for all models are
given. According to Eq.~(\ref{eq:init}) the mass of a binary is on
average 1.5 times the mass of a single star. For each binary in models
A and C we also form a single star, i.e.\ per binary a total of 2.5
times the mass of a single star is formed (1.5 for the binary, 1 for
the single star). For models B and D only 1.5 times the mass of a
single star is formed per binary. Thus for the same SFR in \msun
yr$^{-1}$ the frequency of each process involving a binary of the
models A and C is 0.6 times that in models B and D.

For model A the current birth rate for close double white dwarfs is
$4.8 \, \times \,10^{-2}$ yr$^{-1}$ in the Galaxy. The expected total
population of close binary white dwarfs in the galactic disk is
$\sim\,2.5\,\times\,10^{8}$ (see Table~\ref{tab:birthrates}).

The double white dwarfs are of the following types: 53\% contains two
helium white dwarfs; 25\% two CO white dwarfs; in 14\% a CO white
dwarf is formed first and a helium white dwarf later and in 6\% a
helium white dwarf is formed followed by the formation of a CO white
dwarf. The remaining 1\% of the double white dwarfs contains an ONeMg
white dwarf. The CO white dwarfs can be so called hybrid white dwarfs;
having CO cores and thick helium envelopes \citep{it85,it87}. Of the
double CO white dwarfs, 6\% contains one and 5\% two hybrid white
dwarfs. In the mixed pairs the CO white dwarf is a hybrid in 20\% of
the cases.

Forty eight percent of all systems are close enough to be brought into
contact within a Hubble time.  Most are expected to merge.  The
estimated current merger rate of white dwarfs is
$2.2\,\times\,10^{-2}$ yr$^{-1}$.  The current merger rate of pairs
that have a total mass larger than the Chandrasekhar limit ($M_{\rm
  Ch}$ = 1.44\,\msun) is $3.2\,\times\,10^{-3}$\,yr$^{-1}$.  Since the
merging of binary CO white dwarfs with a combined mass in excess of
$M_{\rm Ch}$\ is a viable model for type Ia SNe \citep[see][for the
most recent review]{liv99}, our model rate can be compared with the
SN~Ia rate of $\sim (4 \pm 1)\,\times\,10^{-3}$\,yr$^{-1}$ for Sbc
type galaxies like our own \citep{cet99}.  In 19\%\ of the systems
that come into contact the ensuing mass transfer is stable and an
interacting double white dwarf (identified with AM CVn stars) is
formed. The model birth rate of AM CVn systems is
$4.6\,\times\,10^{-3}$ yr$^{-1}$ (see Table~\ref{tab:birthrates}).

\subsection{Observed sample of double white dwarfs}\label{obssample}

\begin{table}[t]
\caption[]{Parameters of known close double white dwarfs (first 14
  entries) and subdwarfs with white dwarf companions. $m$ denotes
  the mass of the visible white dwarf or subdwarf. The mass ratio $q$ is
  defined as the mass of the brighter star of the pair over the mass
  of the companion. For references see \citet{mm99,mmm+99,mar99}; and
  \citet{mmm+00}. The mass of 0136$+$768 is corrected for
  a misprint in \citet{mm99}, for 0135$+$052 the
  new mass given in \citet{brl97} is taken. Data for the sdB star
  KPD~0422+5421 are from \citet{ow00} and for KPD~1930+2752 from \citet{mmn00}. The remaining sdB stars do not
  have reliable mass estimates.}
\label{tab:obs}
\begin{center}
\begin{tabular}{lrrr|lr} \hline
WD/sdB & $P$(d) & $q$  & $m
$ & sdB & $P$(d) \\ \hline
0135$-$052 & 1.556 & 0.90 & 0.25 & 0101$+$039 & 0.570  \\
0136$+$768 & 1.407 & 1.31 & 0.44 & 0940$+$068 & 8.33   \\
0957$-$666 & 0.061 & 1.14 & 0.37 & 1101$+$249 & 0.354  \\
1022$+$050 & 1.157 &      & 0.35 & 1432$+$159 & 0.225  \\
1101$+$364 & 0.145 & 0.87 & 0.31 & 1538$+$269 & 2.50   \\
1202$+$608 & 1.493 &      & 0.40 & 2345$+$318 & 0.241  \\
1204$+$450 & 1.603 & 1.00 & 0.51  & &\\
1241$-$010 & 3.347 &      & 0.31 & & \\
1317$+$453 & 4.872 &      & 0.33  & &\\
1704$+$481A& 0.145 & 0.7  & 0.39  & &\\
1713$+$332 & 1.123 &      & 0.38 & & \\
1824$+$040 & 6.266 &      & 0.39  & &\\
2032$+$188 & 5.084 &      & 0.36 & & \\
2331$+$290 & 0.167 &      & 0.39  & & \\[0.1cm] 
KPD~0422+5421 & 0.090 & 0.96  & 0.51 & & \\ 
KPD~1930+2752 & 0.095 & 0.52  & 0.5 & & \\ 
\hline
\end{tabular}
\end{center}
\end{table}

The properties of the observed double white dwarfs with which we will
compare our models are summarised in Table~\ref{tab:obs}. Only
WD~1204+450 and WD~1704+481 are likely to contain CO white dwarfs,
having components with masses higher than 0.46\,\msun; the limiting
mass to form a helium white dwarf \citep{sgr90}. The remaining systems
are probably helium white dwarfs. In principle in the mass range $M
\simeq 0.35 - 0.45\,\msun$\ white dwarfs could also be hybrid; However
in this range the probability for a white dwarf to be hybrid is 4 -- 5
times lower than to be a helium white dwarf, because hybrid white
dwarfs originate from more massive stars which fill their Roche lobe
in a narrow period range \citep[see, however, an example of such a
scenario for WD~0957-666 in][]{nvy+00}. We assume $0.05\,\msun$\ for
the uncertainty in the estimates of the masses of white dwarfs, which
may be somewhat optimistic.

Table~\ref{tab:obs} also includes data on subdwarf B stars with
suspected white dwarf companions. Subdwarf B (sdB) stars are hot,
helium rich objects which are thought to be helium burning remnants of
stars which lost their hydrogen envelope. When their helium burning
has stopped they will become white dwarfs. Of special interest are KPD
0422+5421 \citep{kow98,ow00} and KPD1930+2752 \citep{mmn00}. With
orbital periods as short as 0.09 and 0.095 days, respectively, their
components will inevitably merge. In both systems the sdB components
will become white dwarfs before the stars merge. In KPD1930+2752 the
total mass of the components is close to the Chandrasekhar mass or
even exceeds it. That makes this system the only currently known
candidate progenitor for a SN Ia.

\subsection{Period - mass distribution; constraints on cooling models}
\label{results:P_m}

\begin{figure*}[t]
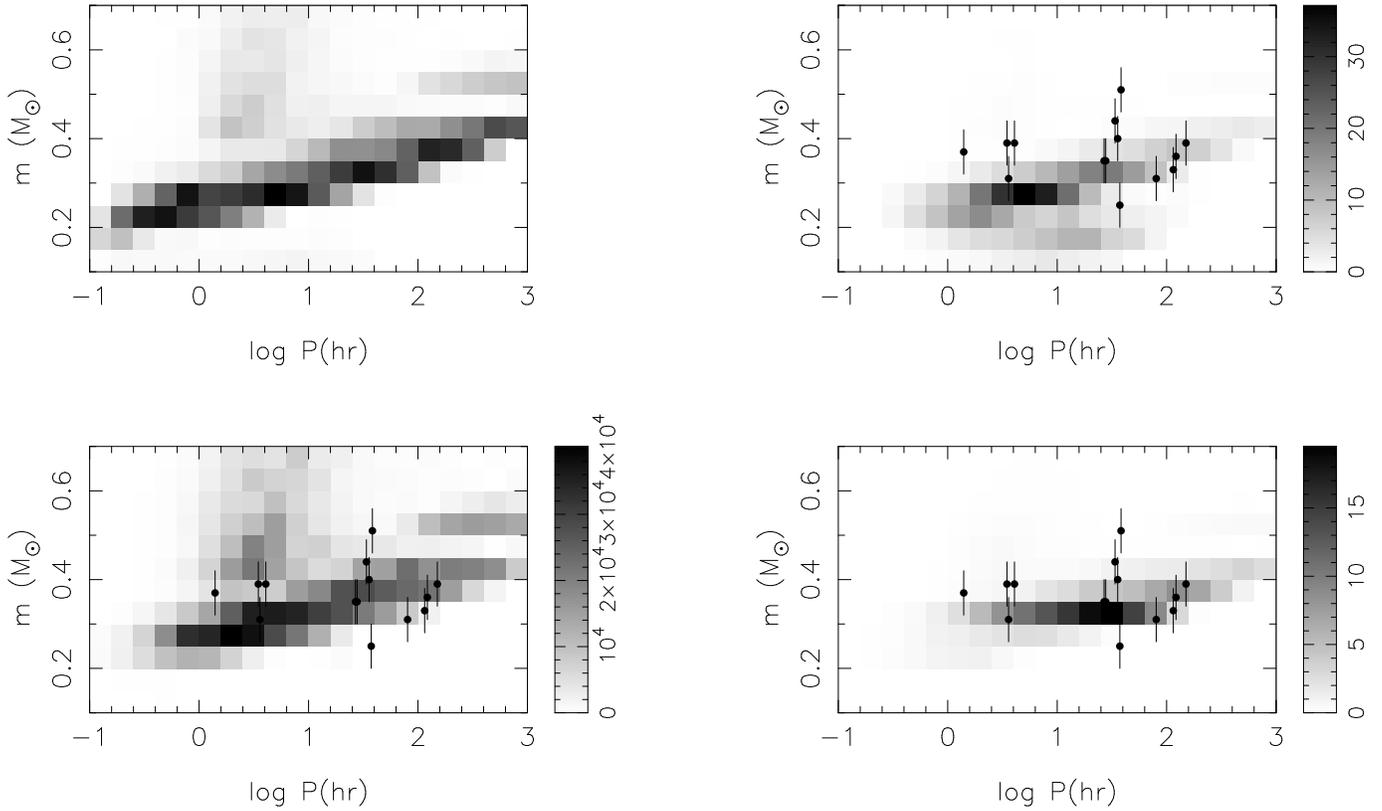

\begin{minipage}{0.45\textwidth}
\psfig{figure=H2305f5.ps,width=\textwidth,angle=-90}
\end{minipage}
\hspace*{0.095\textwidth}
\begin{minipage}{0.45\textwidth}
\psfig{figure=H2305f6.ps,width=\textwidth,angle=-90}
\end{minipage}
\\
\begin{minipage}{0.45\textwidth}
\psfig{figure=H2305f7.ps,width=\textwidth,angle=-90}
\end{minipage}
\hspace*{0.095\textwidth}
\begin{minipage}{0.45\textwidth}
\psfig{figure=H2305f8.ps,width=\textwidth,angle=-90}
\end{minipage}
\caption[]{Model population of double white dwarfs as function of orbital
  period and mass of the brighter white dwarf of the pair. Top left:
  distribution of the double white dwarfs that are currently born for
  models A. This is independent of cooling.  In the remaining three
  plots we show the currently visible population of double white
  dwarfs for different cooling models: (top right) cooling according
  to \citetalias{dsb+98} and \citet[model A1]{blo95b}; (bottom right)
  cooling according to \citetalias{dsb+98}, but with faster cooling
  for WD with masses below 0.3\,\msun\ (model A2). Both plots are for
  a limiting magnitude $V_{\rm lim} =$ 15; (bottom left) with constant
  cooling time of 100 Myr (model A3, note that in this case we only
  obtain the {\em total number of potentially visible double white
    dwarfs in the Galaxy} and we can not construct a magnitude limited
  sample). For comparison, we also plot the observed binary white
  dwarfs. }
\label{fig:P_m}
\end{figure*}

The observed quantities that are determined for all double white
dwarfs are the orbital period and the mass of the brighter white
dwarf. Following \citet{sly98}, we plot in Fig.~\ref{fig:P_m} the
$P_{\rm orb} - m$\ distributions of the frequency of occurrence for
the white dwarfs which are born at this moment and for the simulated
magnitude limited sample for the models with different cooling
prescriptions, (models A1, A2 and A3; see Table~\ref{tab:models}),
where we assume $V_{\rm lim}$ = 15 as the limiting magnitude of the
sample\footnote{The $P - m$\ distribution does not qualitatively
  change if we increase $V_{\rm lim}$ by one or two magnitudes, since
  we still deal with very nearby objects.}. For $m$ we always use the
mass of the brighter white dwarf. In general the brighter white
dwarf is the one that was formed last, but occasionally, it is the one
that was formed first as explained in Sect.~\ref{examples:CO+He}.  For
comparison, we also plot the observed binary white dwarfs in
Fig.~\ref{fig:P_m}.

There is a clear correlation between the mass of new-born low-mass
(He) white dwarf and the orbital period of the pair. This can be
understood as a consequence of the existence of a steep core mass -
radius relation for giants with degenerate helium cores \citep{rw70}.
Giants with more massive cores (forming more massive white dwarfs)
have much larger radii and thus smaller binding energies. To expell
the envelope in the common envelope, less orbital energy has to be
used, leading to a larger orbital period.  The spread in the
distribution is caused by the difference in the masses of the
progenitors and different companion masses.

In the simulated population of binary white dwarfs there are three
distinct groups of stars: He dwarfs with masses below 0.45\,\msun,
hybrid white dwarfs with masses in majority between 0.4 and 0.5\,
\msun\ and periods around a few hours, and CO ones with masses above
0.5\,\msun. The last groups are clearly dominated by the lowest mass
objects. The lowest mass CO white dwarfs are descendants of most
numerous initial binaries with masses of components 1 -- 2\,\msun.

\begin{figure}[t]
\psfig{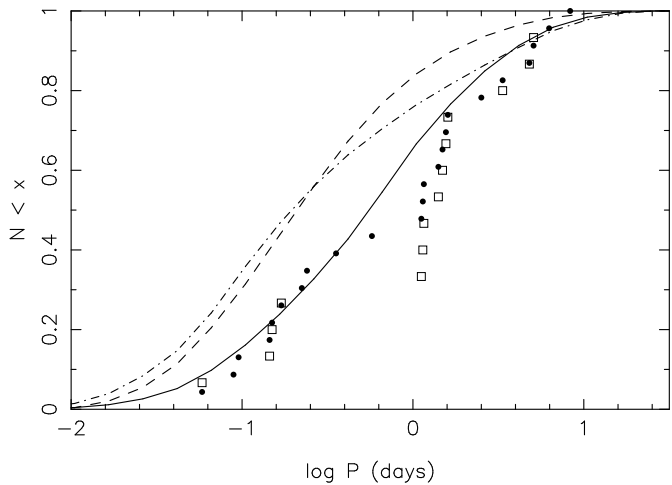}
\caption{Cumulative distribution of periods. Solid line for our best
  model (A2); \citetalias{dsb+98} cooling, but with lower luminosity
  due to thermal flashes for white dwarfs with masses below
  0.3\,\msun.  Dashed line for \citetalias{dsb+98} without
  modifications (model A1) and dash dotted line for constant cooling
  time of 100 Myr (model A3).  Open squares for the observed double
  white dwarfs, filled circles give the observed systems including the
  sdB binaries (Table~\ref{tab:obs}).}
\label{fig:Pcumul}
\end{figure}

The different cooling models result in very different predicted
observable distributions. Model A1 where the cooling curves of
\citetalias{dsb+98} are applied favours low mass white dwarfs to such
an extent that almost all observed white dwarfs are expected to have
masses below 0.3\,\msun.  This is in clear contrast with the
observations, in which all but one white dwarf have a mass above
0.3\,\msun. Reduced cooling times for white dwarfs with masses below
0.3\,\msun\ (model A2) improves this situation.  Model A3, with a
constant cooling time (so essentially only affected by merging due to
GWR), seems to fit all observed systems also nicely.  However, a
complementary comparison with the observations as given by cumulative
distributions of the periods (Fig.~\ref{fig:Pcumul}), shows that model
A2 fits the data best, and that model A3 predicts too many short
period systems.

The observed period distribution for double white dwarfs shows a gap
between 0.5 and 1 day, which is not present in our models. If we
include also sdB binaries, the gap is partially filled in. More
systems must be found to determine whether the gap is real.  

The comparison of our models with observations suggests that white
dwarfs with masses below 0.3\,\msun\ cool faster than predicted by
\citetalias{dsb+98}.  Mass loss in thermal flashes and a stellar wind
may be the cause of this.

The model sample of detectable systems is totally dominated by He
white dwarfs with long cooling times. Given our model birth rates and
the cooling curves we apply, we estimate the number of double white
dwarfs to be detected in a sample limited by $V_{\rm lim}$ = 15 as 220
of which only 10 are CO white dwarfs for model A2. Roughly one of
these is expected to merge within a Hubble time having a total mass
above $M_{\rm Ch}$.  For future observations we give in
Table~\ref{tab:V_lim} a list of expected number of systems for
different limiting magnitudes. 

It should be noted that these numbers are uncertain. This is
illustrated by the range in birth rates for the different models
(Table~\ref{tab:birthrates}) and by the differences with previous
studies (see Sect.~\ref{discussion:birthrates}). Additional
uncertainties are introduced by our limited knowledge of the initial
distributions (Eq.~\ref{eq:init}) and the uncertainties in the cooling
and the Galactic model (Eq.~(\ref{eq:rho})).  For example
\citet{ylt+94} compare models with two different $q_{\rm i}$
distributions (one peaked towards $q_{\rm i} \sim 1$) and show that
the birth rates differ by a factor $\sim1.7$. In general the relative
statistics of the model are more reliable than the absolute
statistics.

\begin{figure}[t]
\psfig{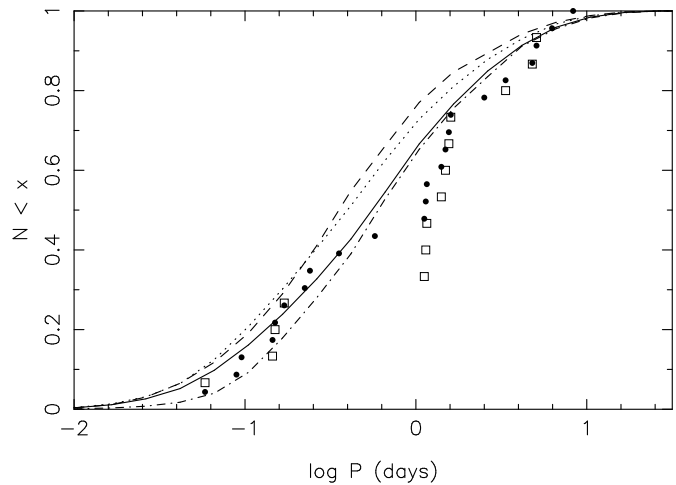}
\caption{Cumulative distribution of periods. Solid line for model A2
  as in Fig.~\ref{fig:Pcumul}, dashed line for the same model but
  with $\alpha_{\rm ce} \lambda$ = 1, dash-dotted line for a model
  with $\gamma$ = 1.5 and finally the dotted line for model C
  (constant SFR).}
\label{fig:Pcumul_parameters}
\end{figure}

\begin{table}[t]
\caption[]{Number of observable white dwarfs, close double white
  dwarfs and SN~Ia progenitors as function of the limiting magnitude
 of the sample for model A2.}
\label{tab:V_lim}
\begin{center}
\begin{tabular}{rrrr}
$V_{\rm lim}$ &  \#wd & \#wdwd & \#SN~Ia prog \\ \hline
15.0 & 855   &  220  &   0.9 \\
15.5 & 1789   &  421  &   1.7 \\
16.0 & 3661   & 789  &   3.2 \\
17.0 &12155   & 2551  &  11.2 \\ \hline
\end{tabular}
\end{center}
\end{table}

Before turning to the mass ratio distribution, we illustrate the
influence of the model parameters we choose. We do this by showing
cumulative period distributions for some models with different
parameters in Fig.~\ref{fig:Pcumul_parameters}; $\alpha_{\rm ce}
\lambda$ = 1 (dashed line) and $\gamma$ = 1.5 (dash-dotted line). It
shows that the change in parameters influences the distributions less
than the different cooling models discussed above, although the
observations favour a higher $\alpha_{\rm ce} \lambda$. We also
included the cumulative distribution for model C (with a constant SFR;
dotted line) which differs from that for model A2 in that it has fewer
long period systems. This is a consequence of the larger relative
importance of old, low-mass progenitor binaries in model A2, which
lose less mass and thus shrink less in the first phase of mass
transfer (see Eq.~\ref{eq:orbit_change}).

\subsection{Period - mass ratio distribution}
\label{sec:p-q}

\begin{figure}[t]
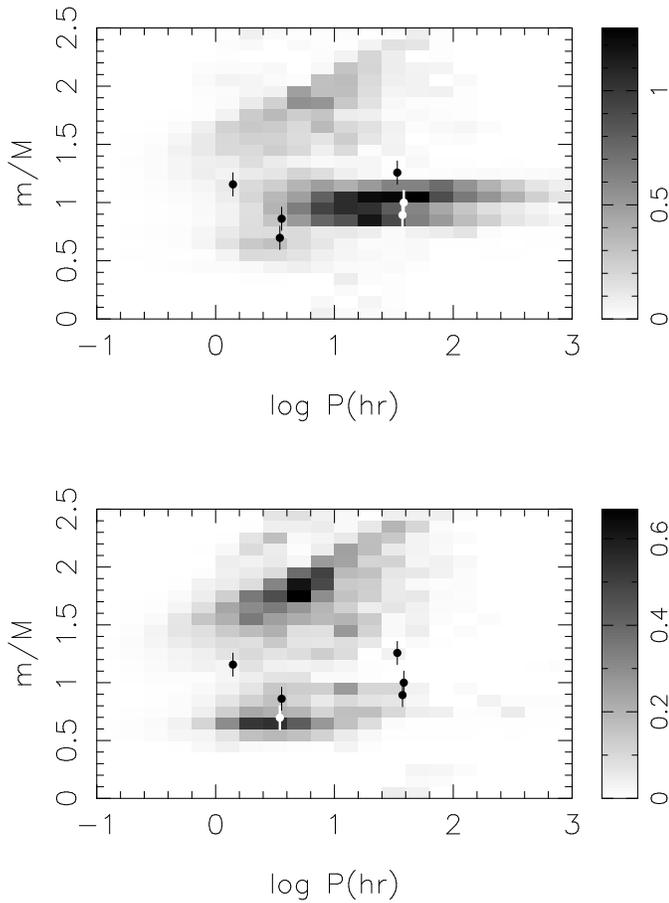

\psfig{figure=H2305f11.ps,width=\columnwidth,angle=-90}
\psfig{figure=H2305f12.ps,width=\columnwidth,angle=-90}
\caption{\textbf{Top:} 
  current population of double white dwarfs as function of orbital
  period and mass ratio, for model A2, a limiting magnitude of 15 and
  a maximal ratio of luminosities of 5.  \textbf{Bottom:} the same for
  a run in which the first phase of mass transfer is treated as a
  standard common envelope, as is done by \citetalias{ity97} and
  \citetalias{han98}. For comparison, we also plot the observed binary
  white dwarfs.}
\label{fig:P_q}
\end{figure}

\begin{figure}[t]
\psfig{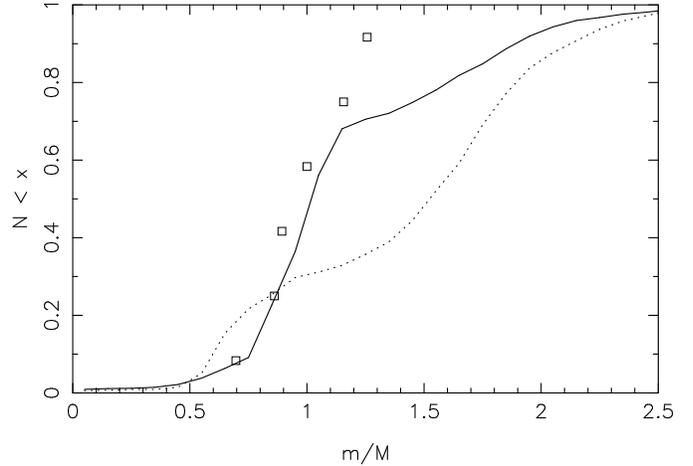}
\caption{ Cumulative mass ratio distributions for the models A2
  (solid line) and A$^\prime$ (dotted line) as explained in
  Sect.~\ref{sec:p-q}. The observed mass ratio's are plotted as the
  open squares.}
\label{fig:qcumul}
\end{figure}

Our assumption that a common envelope can be avoided in the first
phase of mass transfer between a giant and a main-sequence star, is
reflected in the mass ratios of the model systems. A clear prediction
of the model is that close binary white dwarfs must concentrate to $q
= m/M \sim 1.$ For the observed systems, the mass ratio can only be
determined if both components can be seen which in practice requires
that the luminosity of the fainter component is more than 20\% of that
of the brighter component \citep{mmm00}. Applying this selection
criterium to the theoretical model, we obtain the distribution shown
in Fig.~\ref{fig:P_q} for the magnitude limited sample. Note that
since lower mass white dwarfs cool slower this selection criterium
favours systems with mass ratios above unity. In the same figure we
also show the observed systems.
  
For comparison we also computed a run (A$^\prime$) in which we used
the standard common envelope treatment for the first phase of mass
transfer, which is done by \citetalias{ity97} and \citetalias{han98}.
The fraction of double white dwarfs for which the mass ratio can be
determined according to the selection criterium of a luminosity ratio
greater than 0.2, is 27\% for model A2 and 24\% for model A$^\prime$.
In a total of 14 systems one thus expects 4$\pm$2 and 3$\pm$2 systems
of which the mass ratio can be determined. Model A2 fits this number
better, but the numbers are too small to draw conclusions.  The
distribution of mass ratios in model A$^\prime$ (Fig.~\ref{fig:P_q},
bottom) however clearly does not describe the observations as
well as our model A2, as illustrated in more detail in a plot where
the cumulative mass ratio distributions of the two models and the
observations are shown (Fig.~\ref{fig:qcumul}).

\subsection{Mass spectrum of the white dwarf population; constraints
  on the binary fraction}\label{wd_mass}

\begin{figure*}[t]
\begin{minipage}[t]{0.45\textwidth}
\psfig{figure=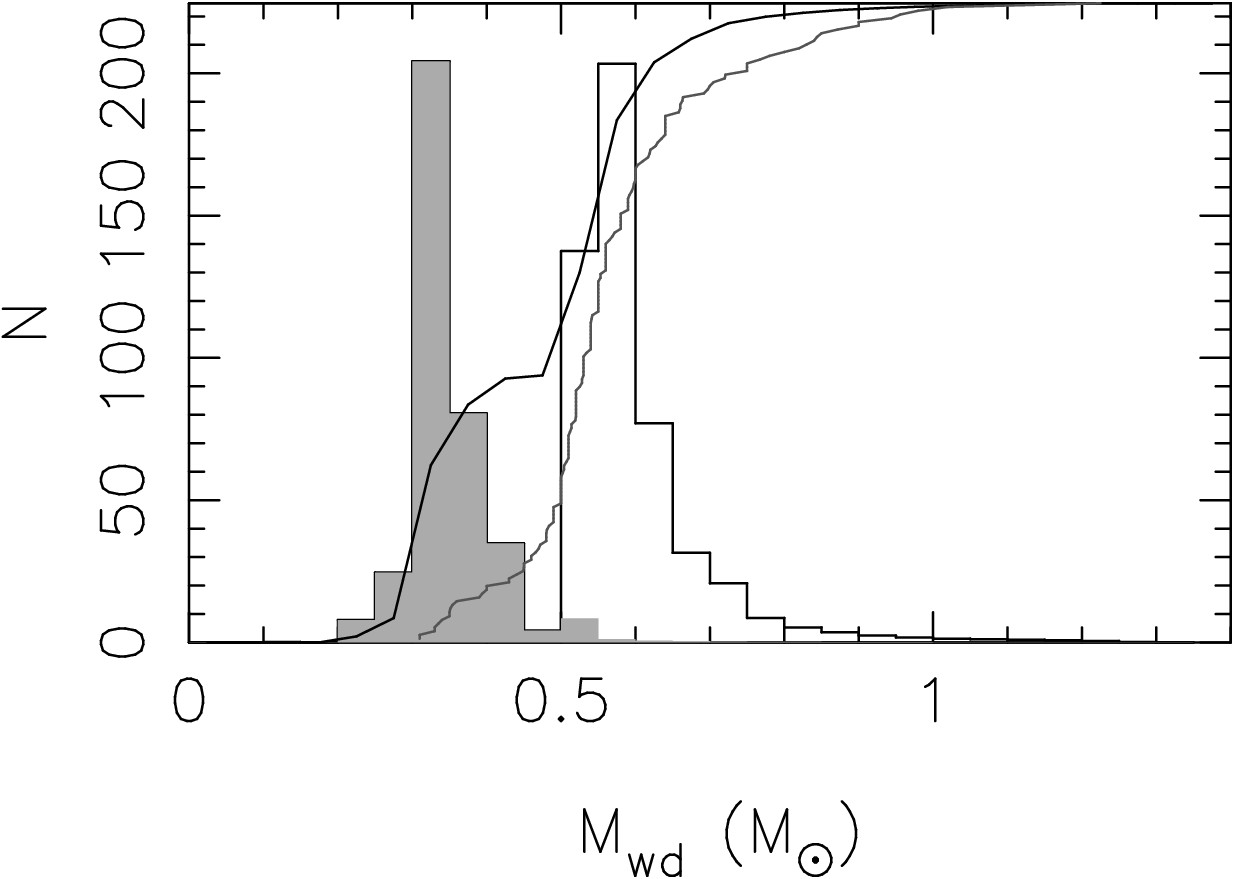,width=\columnwidth,angle=-90}
\caption[]{Mass spectrum of all white dwarfs for model B (100\%
  binaries). Members of close double white dwarfs are in grey. The
  cumulative distribution is shown as the solid black line. For
  comparison, the grey line shows the cumulative distribution of the
  observed systems (Fig.~\ref{fig:wdmass_obs}).  }
\label{fig:wdmass_100}
\end{minipage}
\hspace*{0.095\textwidth}
\begin{minipage}[t]{0.45\textwidth}
\psfig{figure=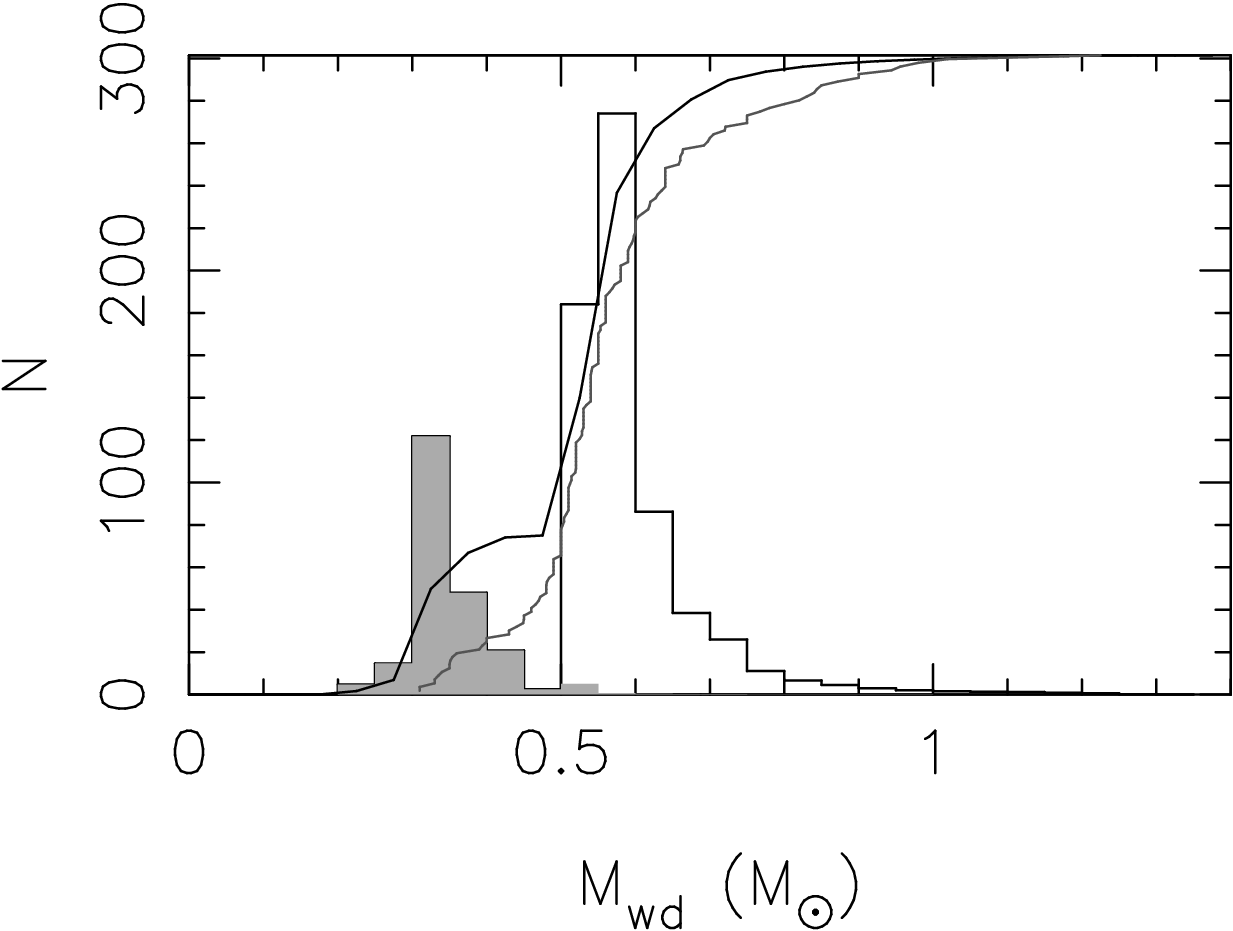,width=\columnwidth,angle=-90}
\caption[]{Mass spectrum of all white dwarfs for model A2 (
  initial binary fraction of 50\%) Double white dwarfs are in grey.
  The cumulative distribution is shown as solid black line and
  cumulative distribution of observed systems as the grey line
}
\label{fig:wdmass_50}
\end{minipage}
\vspace*{0.3cm}\\
\begin{minipage}[t]{0.45\textwidth}
\psfig{figure=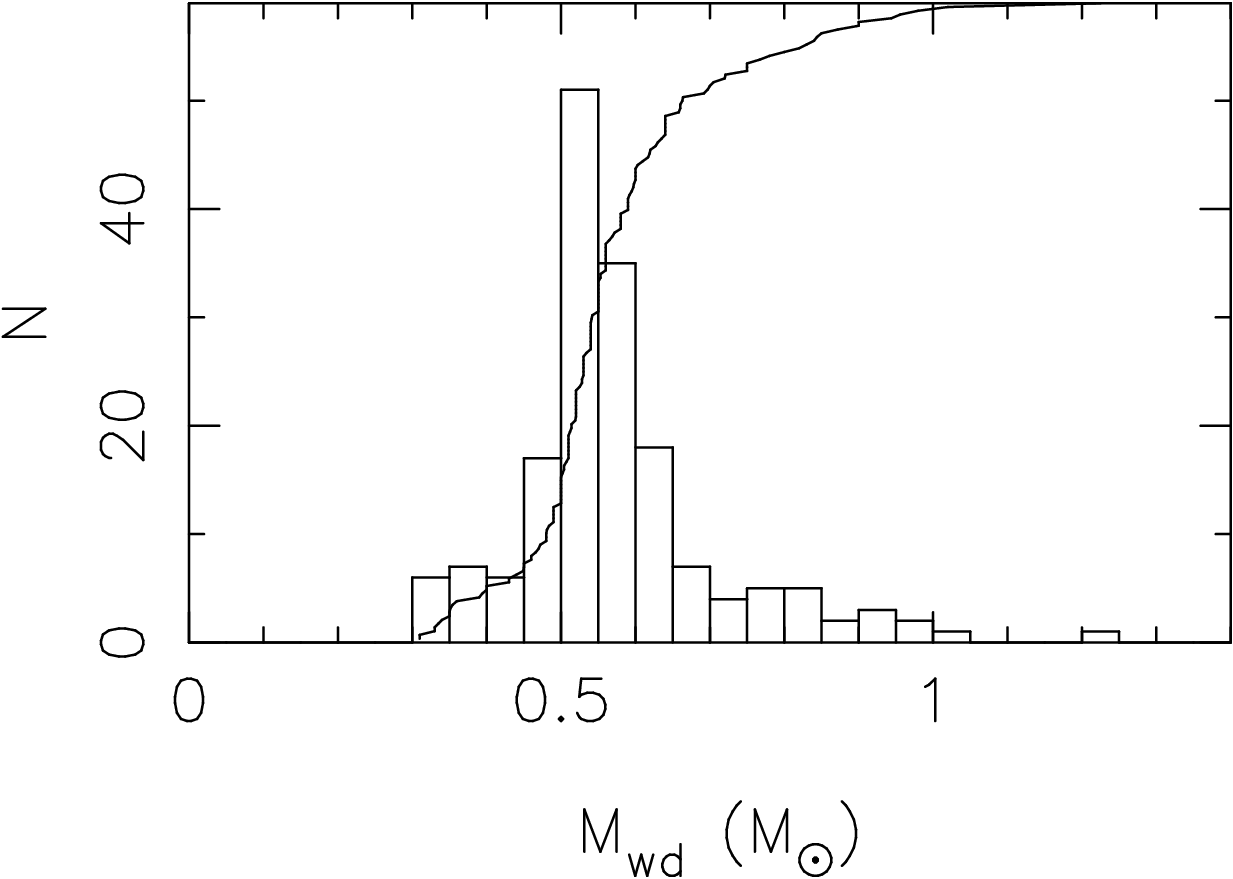,width=\columnwidth,angle=-90}
\caption[]{Mass spectrum of observed white dwarfs. Data are taken from 
  \citet{bsl92} and \citet{brb95}. The solid line is the cumulative
  distribution.}
\label{fig:wdmass_obs}
\vspace*{1.2cm}
\end{minipage}
\hspace*{0.095\textwidth}
\begin{minipage}[t]{0.45\textwidth}
\psfig{figure=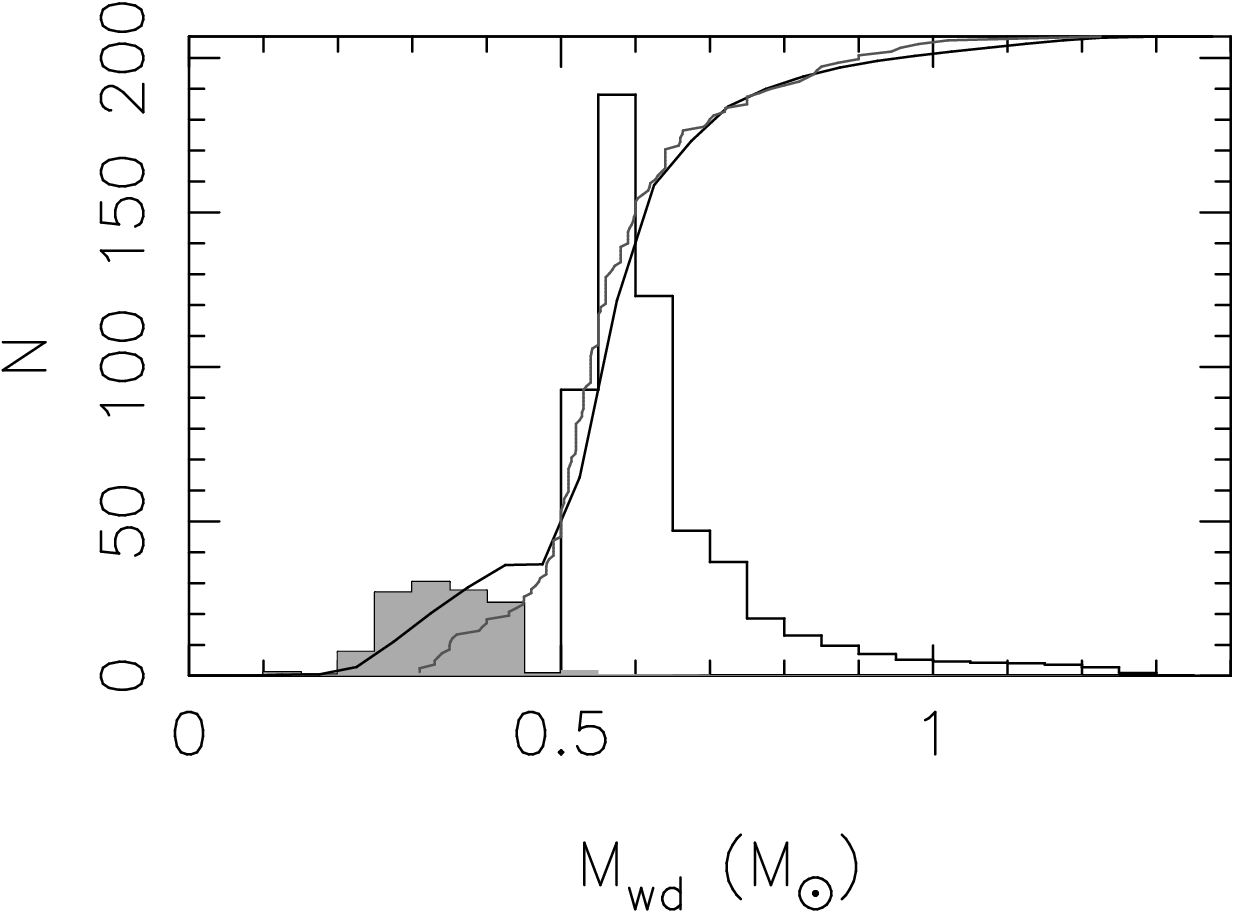,width=\columnwidth,angle=-90}
\caption[]{Mass spectrum of all white dwarfs as in
  Fig.~\ref{fig:wdmass_50} in a model in which all helium white dwarfs
  cool like a 0.4\,\msun\ dwarf and all CO white dwarfs cool like a
  0.6\,\msun\ white dwarf.  Lines are cumulative distributions for the
  model (black) and the observations (grey)
}
\label{fig:wdmass_GN}
\end{minipage}
\end{figure*}

Figs.~\ref{fig:wdmass_100} and \ref{fig:wdmass_50} show the model
spectrum of white dwarf masses for models B and A2, including both
single and double white dwarfs for a limiting magnitude $V_{\rm lim} =
15$. For this plot we consider as ``single'' white dwarfs all objects
that were born in initially wide pairs, single merger products, white
dwarfs that became single as a result of binary disruption by SN
explosions, white dwarfs in close pairs which are brighter than their
main-sequence companions and genuine single white dwarfs for the
models with an initial binary fraction smaller than 100\%.

These model spectra can be compared to the observed mass spectrum of
DA white dwarfs studied by \citet{bsl92} and \citet{brb95}, shown in
Fig.~\ref{fig:wdmass_obs}.  The latter distribution may have to be
shifted to higher masses by about 0.05\,\msun, if one uses models of
white dwarfs with thick hydrogen envelopes for mass estimates
\citep{ngs99}. Clearly, a binary fraction of 50\% fits the observed
sample better, if indeed helium white dwarfs cool much slower than CO
white dwarfs. We can also compare the absolute numbers. \citet{mm99}
conclude that the fraction of close double white dwarfs among DA white
dwarfs is between 1.7 and 19 \% with 95\% confidence. For model B the
fraction of close white dwarfs is $\sim$43\% (853 white dwarfs of
which 368 are close pairs), for model A2 is is $\sim$26\% (855 white
dwarfs and 220 close pairs). Note that this fraction slightly
decreases for higher limiting magnitudes because the single white
dwarfs are more massive and thus generally dimmer, sampling a
different fraction of Galaxy. An even lower binary fraction apparently
would fit the data better, but is in conflict with the estimated
fraction of binaries among normal main sequence binaries
\citep{abt83,dm91}. However this number highly depends on uncertain
selection effects.

There are some features in the model mass spectrum in model A2 that
appear to be in conflict with observations.  The first is the clear
trend that with the cooling models of \citetalias{dsb+98}, even with
our modifications, there should be an increasing number of helium
white dwarfs towards lower masses. The observed distribution is flat.
A very simple numerical experiment in which we assign a cooling curve
to all helium white dwarfs as the one for a 0.414\,\msun\ white dwarf
according to \citetalias{dsb+98} and a cooling curve as for a
0.605\,\msun\ white dwarf according to \citet{blo95b} for all CO white
dwarfs (Fig.~\ref{fig:wdmass_GN}), shows that an equal cooling time
for all helium white dwarfs seems to be in better agreement with the
observations. It has a fraction of double white dwarfs of 18\%.
  
Another feature is the absence of stars with 0.45 $\la M/\msun \la$
0.5 in the model distributions. This is a consequence of the fact that
in this interval in our models only hybrid white dwarfs can be
present, which have a low formation probability (see
Sect.~\ref{obssample}).

We conclude that an initial binary fraction of 50\% can explain the
observed close binary fraction in the white dwarf population. The
shape of the mass spectrum, especially for the helium white dwarfs is
a challenge for detailed mass determinations and cooling models.

\subsection{Birth rate of PN and local WD space density; constraints on 
the star formation history}\label{PN}

\begin{table}[t]
\caption{Galactic number and local space density of white
  dwarfs; and Galactic and local PN formation
  rate for the models A and C. Unit of the PN formation rates
  is yr$^{-1}$; unit for $\rho_{\rm wd,\sun}$ is pc$^{-3}$.
  The ranges of observed values are given for comparison. For
  references and discussion see Sect.~\ref{PN}}
\label{tab:space_densities}
\begin{center}
\begin{tabular}{lrrrrrr}
Model & SFH    & \% bin& \#wd& $\nu_{\rm PN}$ & $\rho_{\rm wd,\sun}$ & 
$\nu_{\rm PN, \sun}$  \\
      &        &     & $10^9$&     &$(10^{-3})$& $(10^{-12})$  \\ \hline
A     & Exp    & 50  &   9.2 &   1.1 & 19~~~ &  2.3  \\  
C     & Const  & 50  &   4.1 &   0.8 &  8.5 &  1.7  \\
Obs   &        &     &       &        & 4-20  &    3  \\ \hline
\end{tabular}
\end{center}
\end{table}

Finally, we compare models A and C (see Table~\ref{tab:models}),
which differ only by the assumed star formation history. The star
formation rate was probably higher in the past than at present and
some (double) white dwarfs descend from stars that are formed just
after the galactic disk was formed.

Table~\ref{tab:space_densities} gives the formation rates of PN and
the total number of white dwarfs in the Galaxy for models A and C.
The total number of white dwarfs is computed by excluding all white
dwarfs in binaries where the companion is brighter.
The local density of white dwarfs and PN rate are computed with
Eq.~(\ref{eq:rho_sun}) as described in Sect.~\ref{rho}.

We can compare these numbers with the observational estimates for the
local PN formation rate of 3\,$\times\,10^{-12}$ pc$^{-3}$ yr$^{-1}$
\citep{pot96} and the local space density of white dwarfs, which range
from e.g.  $4.2\,\times\,10^{-3}$ pc$^{-3}$\ \citep{khh99} through
$7.6^{+3.7}_{-0.7}\,\times\,10^{-3}$ pc$^{-3}$\ \citep{osw+95} and
$10\,\times\,10^{-3}$ pc$^{-3}$\ \citep{rt95} to $20 \pm
7\,\times\,10^{-3}$ pc$^{-3}$\ \citep{fes98}.

This list shows the large uncertainty in the observed local space
density of white dwarfs. It appears that the lower values are somewhat
favoured in the literature. Both models A and C appear for the moment
to be consistent with the observed local white dwarf space density and
with the PN formation rate. However, we prefer model A2 since it fits
the period distribution better (see Fig.~\ref{fig:Pcumul_parameters}).

The ratio of the local space density of white dwarfs to the current
local PN formation rate could in principle serve as a diagnostic for
the star formation history of the Galaxy, given better knowledge of
$\rho_{\rm wd,\sun}$, which critically depends on the estimates of the
incompleteness of the observed white dwarf samples and the applied
cooling curves.

\section{Discussion: comparison with previous studies}\label{discussion}

We now compare our work with the results of previous studies; in
particular the most recent studies of \citet[\citetalias{ity97}]{ity97} 
and \citet[\citetalias{han98}]{han98}.

\subsection{Birth rates}\label{discussion:birthrates}

In Table~\ref{tab:birthrates} we show the birth rates of close double
white dwarfs for the different models. We also include numbers from
\citetalias{han98} (model 1) and a set of numbers computed with the
same code as used in \citetalias{ity97}, but for an age of the
galactic disk of 10 Gyr, as our models.  The numbers of
\citetalias{han98} are for an age of the disk of 15 Gyr.  Our model D
is the closest to the models of \citetalias{ity97} and
\citetalias{han98}, assuming a constant SFR and 100\% binaries. To
estimate the influence of the binary evolution models only in
comparing the different models we correct for their different
normalisations.

In the recomputed \citetalias{ity97} model the formation rate of
interacting binaries in which the primary evolves within the age of
the Galaxy is 0.35 yr$^{-1}$. In our model D this number is 0.25
yr$^{-1}$. In the following we therefore multiply the formation rates
of \citetalias{ity97} as given in Table~\ref{tab:birthrates} with
0.71.

In the model of \citetalias{han98} one binary with a primary mass
above 0.8 \msun\ is formed in the Galaxy annually with $\log a_{\rm i}
< 6.76$, i.e.\ 0.9 binary with $\log a_{\rm i} < 6$, which is our
limit to $a_{\rm i}$. Correcting for the different assumed age of the
Galaxy we estimate this number to be 0.81; in our model this number is
0.73.  We thus multiply the the formation rates of \citetalias{han98}
as given in Table~\ref{tab:birthrates} with 0.9.

Applying these corrections to the normalisation, we find that some
interesting differences remain. The birth rate of double white dwarfs
is 0.029, 0.053 and 0.062 per year for \citetalias{han98}, model D and
ITY98 respectively. At the same time the ratio of the merger rate to
the birth rate decreases: 0.97, 0.53 and 0.28 for these models. This
can probably be attributed to the different treatment of the common
envelope.  \citetalias{han98} uses a common envelope spiral-in
efficiency of 1 in Webbinks \citeyearpar{web84} formalism, while we
use 4 \citep[for $\lambda = 0.5$, see][]{khp87}.  \citetalias{ity97}
use the formalism proposed by \citet{ty79} with an efficiency of 1.
This is comparable to an efficiency of 4 -- 8 in the Webbink
formalism. This means that in the model of \citetalias{han98}, more
systems merge in a common envelope, which yields a low formation rate
of double white dwarfs.  The ones that form (in general) have short
periods for the same reason, so the ratio of merger to birth rate is
high. In the ITY model the efficiency is higher, so more systems will
survive both common envelopes and have generally wider orbits, leading
to a much lower ratio of merger to birth rate. Our model D is somewhat
in between, but also has the different treatment of the first mass
transfer phase (Sect~\ref{binev}), in which a strong spiral-in is
avoided.

The difference between the models in the SN~Ia rate ($\nu_{\rm
  SN~Ia}$) is related both to the total merger rate and to the
\textit{masses} of the white dwarfs.  The former varies within a
factor $\sim 1.5$: 0.017, 0.028, and 0.028 ${\rm yr}^{-1}$\ for
\citetalias{ity97}, model D, and \citetalias{han98}, while $\nu_{\rm
  SN~Ia}$ is higher by a factor 2 -- 3 in model D compared to the
other models.  This is caused by the initial - final mass relation in
our models, which is derived from stellar models with core
overshooting, producing higher final masses.

The difference in the birthrate of interacting white dwarfs ($\nu_{\rm
  AM CVn}$) is mainly a consequence of our treatment of the first mass
transfer, which gives for model D a mass ratio distribution which is
peaked to 1 (Sect.~\ref{sec:p-q}), while in \citetalias{ity97} and
\citetalias{han98} the mass ratio is in general different from 1
(Sect.~\ref{discussion:Pmq}), favouring stable mass transfer and the
formation of AM CVn systems.  An additional factor, which reduces the
number of AM~CVn systems is the assumption in model D and
\citetalias{ity97} that the mass transfer rate is limited by the
Eddington rate. The formation and evolution of AM~CVn stars is
discussed in more detail in \citet{ty96} and Nelemans et al. (in 
preparation).

\subsection{Periods, masses and mass ratios}\label{discussion:Pmq}

Comparing our Fig.~\ref{fig:P_m} with the corresponding Figure in
\citet{sly98}, we find the same trend of higher white dwarf masses at
longer periods.  However, in our model the masses are higher than in
the model of \citet{sly98} at the same period.  This is a consequence
of the absence of a strong spiral-in in the first mass transfer phase
in our model, contrary to the conventional common envelope model, as
discussed in \ref{masstransfer}.

In our model the mass ratio distribution is peaked at $q \approx 1$.
This is different from the models of \citetalias{ity97} and
\citet{sly98} which predict a strong concentration to $q \sim 0.5 -
0.7$ and from \citetalias{han98} who finds typical values of $q \sim
0.5$, with a tail to $q \sim 2$.  The difference between these two
latter groups of models may be understood as a consequence of enhanced
wind in Han's model \citep[see also][]{te88}, which allows wider
separations before the second common envelope. The mass ratio
distribution of our model, peaked at $q \simeq 1$, appears to be more
consistent with the observed mass ratio distribution.

\subsection{Cooling}\label{discussion:cooling}

To explain the lack of observed white dwarfs with masses below 0.3
\msun\ we had to assume that these white dwarfs cool faster than
predicted by the models of \citetalias{dsb+98}.

The same assumption was required by \citet{kbk+00}, to
bring the cooling age of the white dwarf that accompanies PSR B1855+09
into agreement with the pulsar spin-down age; and to obtain cooling
ages shorter than the age of the Galaxy for the white dwarfs
accompanying PSR J0034$-$0534 and PSR J1713+0747 .

The absence of the lowest mass white dwarfs may also be explained by
the fact that a common envelope involving a giant with a low mass
helium core ($M_{\rm c} < 0.2 - 0.25 \msun$) always leads to a
complete merger, according to \citet{stb00}. However it can not
explain the absence of the systems with $0.25 < M < 0.3 \msun$, which
would form the majority of the observed systems using the full
\citetalias{dsb+98} cooling (model A1; see Fig.~\ref{fig:P_m}).

\section{Conclusions}\label{conclusion}

We computed a model of the population of close binary white dwarfs and
found good agreement between our model and the observed double white
dwarf sample. A better agreement with observations compared to earlier
studies is found due to two modifications.

The first is a different treatment of unstable mass transfer from a
giant to a main sequence star of comparable mass.  The second is a
more detailed modelling of the cooling of low mass white dwarfs which
became possible because detailed evolutionary models for such white
dwarfs became available. Our main conclusions can be summarised as
follows.

\begin{enumerate}
\item Comparing the mass distribution of the white dwarfs in close
  pairs with the observations, we find a lack of observed white dwarfs
  with masses below 0.3 \msun. This discrepancy can be removed with
  the assumption that low-mass white dwarfs cool faster than computed
  by \citet{dsb+98}.  The same assumption removes discrepancies
  between observed and derived ages of low-mass white dwarfs that
  accompany recycled pulsars, as shown by \citet{kbk+00}. Faster
  cooling is expected if the hydrogen envelopes around low-mass white
  dwarfs are partially expelled by thermal flashes or a stellar wind.

\item Our models predict that the distribution of mass ratios of
  double white dwarfs, when corrected for observational selection
  effects as described by \citet{mmm00}, peaks at a mass ratio of
  unity, consistent with observations. The distributions predicted in
  the models by \citet{ity97} and \citet{han98} peak at mass ratios of
  about 0.7 and above 1.5 and agree worse with the observations even
  after applying selection effects.
    
\item Our models predict a distribution of orbital periods and masses
  of close double white dwarfs in satisfactory agreement with the
  observed distribution.

\item Amongst the observed white dwarfs only a small fraction are
  members of a close pair. To bring our models into agreement with
  this, we have to assume an initial binary fraction of 50\%\ (i.e.\ 
  as many single stars as binaries).

\item In our models the ratio of the local number density of white
  dwarfs and the planetary nebula formation rate is a sensitive
  function of the star formation history of the Galaxy. Our predicted
  numbers are consistent with the observations.

  \item Using detailed cooling models we pridict that an observed
    sample of white dwarfs near the Sun, limited at the magnitude
    $V=15$, contains 855 white dwarfs of which 220 are close pairs.
    Of these pairs only 10 are double CO white dwarfs and only one is
    expected to merge having a combined mass above the Chandrasekhar
    mass. The predicted merger rate in the Galaxy of double white
    dwarfs with a mass that exceeds the Chandrasekhar mass is
    consistent with the inferred SN~Ia rate.
    
    \citetalias{ity97} estimated, depending on $\alpha_{\rm ce}$, to
    find one such pair in a sample of $\sim$200 to $\sim$600 white
    dwarfs.  Reversing this argument, when the statistics become more
    reliable, the observed number of systems with different types of
    white dwarfs could provide constraints on the cooling models for
    these white dwarfs.

\end{enumerate}

\begin{acknowledgements}
  We thank the referee A. Gould for valuable comments.  LRY and
  SPZ acknowledge the warm hospitality of the Astronomical Institute
  ``Anton Pannekoek''. This work was supported by NWO Spinoza grant
  08-0 to E.~P.~J.~van den Heuvel, the Russian Federal Program
  ``Astronomy'' and RFBR grant 99-02-16037 and by NASA through Hubble
  Fellowship grant HF-01112.01-98A awarded (to SPZ) by the Space
  Telescope Science Institute, which is operated by the Association of
  Universities for Research in Astronomy, Inc., for NASA under
  contract NAS\, 5-26555.
\end{acknowledgements}

\bibliography{journals,binaries}
\bibliographystyle{NBaa}

\newpage
\appendix
\section{Population synthesis code \textsf{SeBa}}\label{A}

We present some changes we made to the population synthesis code
\textsf{SeBa} \citep[see][]{pv96,py98}.

\subsection{Stellar evolution}

As before, the treatment of stellar evolution in our code is based on
the fits to detailed stellar evolutionary models \citep{eft89,tap+97},
which give the luminosity and the radius of the stars as a function of
time and mass. In addition to this we need the mass of the core and 
the mass loss due to stellar wind. These we obtain as follows.

\subsubsection{Core mases and white dwarf masses}\label{a:wdmass}

For the mass of the helium core $m_{\rm c }$ at the end of the main
sequence we use (Eggleton, private communication, 1998)
\begin{equation}
m_{\rm c } =  \frac{0.11 \; M^{1.2} + 7\,\times\,10^{-5} \; M^4}{ 1 + 2\,\times\,10^{-4} \, M^3}.
\end{equation}
The mass of the core during the further evolution of the star is
computed by integrating the growth of the core resulting from
hydrogen shell burning:
\begin{equation}
\dot{m_{\rm c}} = \eta_H \, \frac{L}{X}
\end{equation}
where
\begin{equation}
\eta_H = 9.6 \,\times\,10^{-12} \msun \; {\rm yr}^{-1} \; {\rm L}_{\odot}^{-1}
\end{equation}
and $X$ is the mass fraction of hydrogen in the envelope. During core
helium burning we assume that half of the luminosity of the star is
produced by hydrogen shell burning, while in the double shell burning
phase we assume that all of the luminosity is produced by the hydrogen
shell burning.

When giants have degenerate cores, application of a core mass -
luminosity relation gives more accurate results than direct
integration of the growth of the core.

For degenerate helium cores of stars with $M \la 2.3 \msun$ we use
\citep{bs88}
\begin{equation}
M_{\rm c} = 0.146 \; L^{0.143}
\label{eq:mlhe}
\end{equation}
(all quantities in solar units).  For degenerate CO cores of stars
with $M \la 8 \msun$ on the AGB we use \citep{gj93}
 \begin{eqnarray}
M_{\rm c} & = & 0.015 + \sqrt{\frac{L}{47488} + 0.1804}  \qquad \qquad L
< 15725 \nonumber \\
M_{\rm c} & = & 0.46 + \frac{L}{46818} \; M^{-0.25}  \qquad \qquad L > 15725
\label{eq:mlco}
\end{eqnarray}
where the transition between the two fits occurs at $M_{\rm c} \approx
0.73 \msun$\ in stars $\sim 3.5 \msun$\ where the two relations fit
together reasonably. We changed the power of the dependence on $M$
from -0.19 in the original paper to -0.25 because the maximum
luminosities given by our fits otherwise lead to white dwarf masses
too high compared to initial - final mass relations as found from
observations \citep[see][]{gj93}.

The masses of CO cores formed by central He burning inside the helium
core are defined in the same way as we define the relation between the
mass of helium stars and their CO cores (see
Sect.~\ref{a:helium_stars}).

\begin{figure}[t]
\psfig{figure=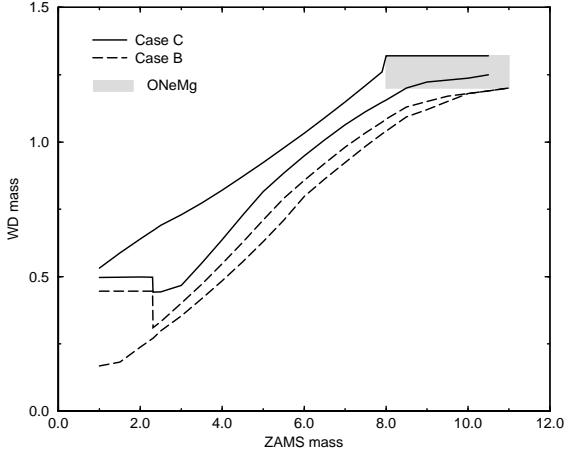,width=\columnwidth,angle=-90}
\caption[]{White dwarf masses as function of the ZAMS mass. Dashed
  lines are for case B mass transfer. The white dwarfs that descend
  from stars with ZAMS masses below 2.3 $\msun$\ in case B mass
  transfer are helium white dwarfs. The two dashed lines give the
  minimum and maximum mass of the white dwarf, which depends on the
  orbital separation at the onset of the mass transfer.  Solid lines
  are for case C mass transfer, which results in the formation of a CO
  white dwarf. When the ZAMS mass is above 8 \msun\ the stripping of
  the envelope in case C mass transfer may prevent the formation of a
  neutron star, leading to the formation of a white dwarfs with a core
  consisting of O, Ne and Mg (shaded region).}
\label{fig:wd_masses}
\end{figure}

A white dwarf forms if a component of a binary with $M < 10 \msun$
loses its hydrogen envelope through RLOF either before core helium
burning (case B mass transfer) or after helium exhaustion (case C).
The masses of white dwarfs formed in cases B and C as function of initial
mass are shown in Fig.~\ref{fig:wd_masses}.

\subsubsection{Helium stars}\label{a:helium_stars}

A helium star is formed when a star more massive than 2.3\,\msun\ 
loses its hydrogen envelope in case B mass transfer. The helium star
starts core helium burning and forms a CO core.  In our code, this
core grows linearly at a rate given by the ratio of 65\% of the
initial mass of the helium star and the total lifetime of the helium
star. This is suggested by computations of \citet{hab86} and gives a
CO core of the Chandrasekhar mass for a 2.2 \msun helium star; the
minimum mass to form a neutron star in our code.

Helium stars with $ 0.8 \la M \la 3 \msun$\ expand again after core
helium exhaustion and can lose their remaining helium envelope in so
called case BB mass transfer. The amount of mass that can be lost is
defined as increasing linearly from 0 to 45\% for stars between 0.8
and 2.2 $\msun$\ and stays constant above 2.2 \msun. The maximum
mass of the CO white dwarf thus formed is 1.21 $\msun$.  Helium stars
of lower mass ($ M < 0.8 \msun$) do not expand and retain their thick
helium envelopes, forming hybrid white dwarfs \citep{it85}.

\subsubsection{Stellar wind}\label{a:stellar_wind}

We describe mass loss in a stellar wind in a very general way in
which the amount of wind loss increases in time according to
\begin{equation}
\Delta M_{\rm w} = M_{\rm lost} \; \left[ \left( \frac{t + \Delta t}{t_{\rm f}} 
\right)^\eta - \left( \frac{t}{t_{\rm f}} \right)^\eta  \right].
\label{eq:sw}
\end{equation}
The exponent $\eta = 6.8$\ is derived from fitting stellar wind mass
loss on the main sequence of massive stars \citep[$M \ga 15 \msun
$][]{mms+94}, but we apply it also for low and intermediate mass
stars. For these stars $t_{\rm f}$ is the duration of the evolutionary
phase that the star is in \citep[as given by][]{eft89}. For the
different evolutionary phases, the parameters $M_{\rm lost}$ is
defined as follows.

In the Hertzsprung gap $M_{\rm lost}$ is 1\% of the total mass of the
star.

For the first giant branch (hydrogen shell burning), we use a fit to
models of \citet{sgr90} for stars with degenerate helium cores
\begin{equation}
M_{\rm lost} = (2.5 - M)/7.5 \quad \msun
\end{equation}
which we extend to all low and intermediate mass stars by setting
$M_{\rm lost} = 0$ above $M = 2.5 \msun$. 

On the horizontal branch  $M_{\rm lost}$ is 5\% of the envelope mass.

For the AGB phase we take $M_{\rm lost}$ equal to 80\% of the mass of
the envelope of the star when it enters the early AGB phase.

\subsubsection{Radii of gyration}

In the previous version of the \textsf{SeBa} code all gyration radii
were set to 0.4. The gyration radius plays a role in the determination 
of the stability of the mass transfer \citep[Appendix C.1]{pv96}. We
now use the following values.

For main-sequence stars we use a fit to the results by \citet{cg90}.
Further we classify stars either as radiative (stars in Hertzsprung
gap and helium stars) or as convective (red giants, AGB stars). A
summary of radii of gyration are given in Table~\ref{tab:gyration}.

\begin{table}[ht]
\caption[]{Gyration radii for various types of stars}
\label{tab:gyration}
\begin{center}
\begin{tabular}{p{0.5\columnwidth}r}
Type & $k^2$  \\
\hline
Radiative stars & 0.03 \\
Convective stars & 0.2 \\
White dwarfs  & 0.4 \\ 
Neutron stars  & 0.25$^a$     \\ 
Black holes  & $\frac{\displaystyle 1}{\displaystyle c \; R^2}$    \\ \hline
\end{tabular}\\
(a) \citet{go69}\\
\end{center}
\end{table}

\subsubsection{White dwarf evolution: luminosity and
radius}\label{a:wd_cooling}

We model the cooling of white dwarfs according to the results of
\citet{blo95b} and \citet{dsb+98}.

\begin{figure}[t]
\psfig{figure=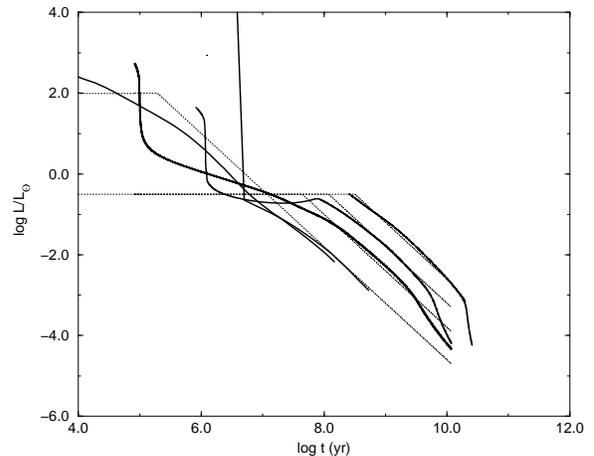,width=\columnwidth,angle=-90}
\caption[]{White dwarf cooling tracks from \citet{dsb+98} and
\citet{blo95b}. Straight lines are the fits to these
curves. The curves are for masses of 0.179, 0.300, 0.414, 0.6 and 0.8
from top right to bottom left}
\label{fig:wdcooling}
\end{figure}

\vspace*{0.3cm}
\noindent {\em Luminosity}\\
The luminosity of white dwarfs as function of time $t$\ can be reasonably
well modelled by
\begin{equation}\label{eq:Lwd}
\log L = L_{\rm max} - 1.4 \log (t/10^6 {\rm yr})
\end{equation}
where $L_{\rm max}$ is a linear fit given by
\begin{equation}\label{eq:Lwd_max}
L_{\rm max} = 3.83 - 4.77 \; M_{\rm WD} \quad \mbox{ for } 
                    0.18 < M_{\rm WD} < 0.6
\end{equation}
(mass and luminosity in solar units).  Outside these limits $L_{\rm
  max}$ stays constant (i.e. $L_{\rm max} = 3$ below $M_{\rm WD} =
0.18$ and $L_{\rm max} = 1$ above $M_{\rm WD} = 0.6$). For white dwarf
masses below $0.6 \msun$\ the luminosity is constrained to be below
$\log L/L_{\sun} = -0.5$, for more massive white dwarfs below $\log
L/L_{\sun} = 2$. In Fig.~\ref{fig:wdcooling} we show the fits and the
results of \citet{blo95b} and \citet{dsb+98}.

\begin{figure}[t]
\psfig{figure=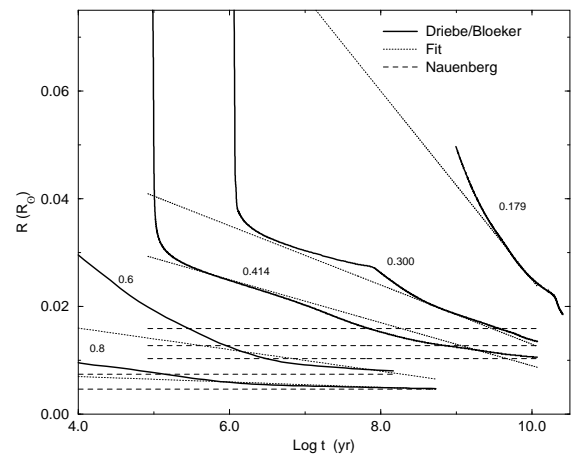,width=\columnwidth,angle=-90}
\caption[]{White dwarf radii from \citet{dsb+98} and
\citet{blo95b}. Straight lines are the fits to these curves. The
curves are for masses of 0.179, 0.300, 0.414, 0.6 and 0.8 from top
right to bottom left}
\label{fig:wdradius}
\end{figure}

\vspace*{0.3cm}
\noindent {\em Radius}\\
We fitted the models of \citet{dsb+98} and \citet{blo95b}, and
interpolated between the fits. The fits are given by
\begin{equation}\label{eq:Rwd}
\frac{R}{R_\odot} = a - b \log (t/10^6 {\rm yr}) \quad \mbox{ for } 
                    M_{\rm WD} < 0.6\,\msun.
\end{equation}
The coefficients $a$ and $b$ are given in
Table~\ref{tab:coefficients}. Fig.~\ref{fig:wdradius} shows the fits
and the corresponding detailed calculations.

For more massive white dwarfs we use the mass-radius relation for
zero-temperature spheres \citep{nau72}
\begin{equation}\label{eq:Rnauenberg}
\frac{R}{R_\odot} = 0.01125 \; \sqrt{\left ( \frac{M_{\rm WD}}{M_{\rm Ch}} \right
)^{-2/3} 
- \left ( \frac{M_{\rm WD}}{M_{\rm Ch}} \right )^{2/3}}
\end{equation}

\begin{table}[ht]
\caption[]{Coefficients for the fits to the white dwarf radii}
\label{tab:coefficients}
\begin{center}
\begin{tabular}{lrr}
$M_{\rm WD}$ & $a$   & $b$ \\ \hline
0.2          & 0.1   & 0.0175 \\
0.4          & 0.03  & 0.0044  \\
0.6          & 0.017 & 0.001   \\
0.8          & 0.011 & 0.0005  \\ \hline
\end{tabular}
\end{center}
\end{table}

\subsubsection{Modified \citetalias{dsb+98} cooling}\label{a:driebe+}

Our modification to the cooling described above reduces the cooling
time scale for white dwarfs with masses below $0.3 \msun$. For these
white dwarfs we use the cooling curve and the radius of a more
massive, thus faster cooling white dwarf of 0.46 \msun\ (see
Sect.~\ref{cooling}).

\subsection{Mass transfer in binary stars}

As suggested by \citet{nvy+00}, we distinguish four types of mass
transfer with different outcomes: stable mass transfer, common
envelope evolution, envelope ejection and a double spiral-in.

\subsubsection{Stable mass transfer}\label{a:stable}

The amount of mass that can be accreted by a star is limited by its
thermal time scale
\begin{equation}
\dot{M}_{\rm max} \approx \frac{M}{\tau_{\rm th}} \approx \frac{R \; L}{G \; M}\,.
\end{equation}
If not all mass can be accreted, we assume that the excess of
mass leaves the system taking with it $n_{\rm J}$\ times the specific
angular momentum of the binary.

This assumption gives for the variation of orbital separation
\begin{equation}
\frac{a_{\rm f}}{a_{\rm i}} = \left(\frac{M_{\rm f} \; m_{\rm
f}}{M_{\rm i} \; m_{\rm i}}\right)^{-2} \;
\left(\frac{M_{\rm f} + m_{\rm f}}{M_{\rm i} + m_{\rm i}}\right)^{2 n_{\rm J} + 1}
\end{equation}
We use $n_{\rm J} = 2.5$, which gives good agreement for the periods
of low-mass Algols and Be X-ray binaries \citep{por96}

\subsubsection{Standard common envelope}\label{a:stace}

When the mass transfer is unstable due to a tidal instability, the
accretor is a compact object, or the envelope ejection equation gives
a smaller orbital separation, we apply the standard common envelope
equation $E_{\rm bind} = \alpha_{\rm ce} \; \Delta E_{\rm orb}$
\citep{web84}:
\begin{equation}\label{eq:ce}     
\frac{M_{\rm i} \; (M_{\rm i} - M_{\rm f})}{\lambda \; R}  
 = \alpha_{\rm ce} \; \left[ \frac{M_{\rm f}\; m}{2 \;a_{\rm f}} -
\frac{M_{\rm i} \; m}{2 \;a_{\rm i}} \right]
\end{equation}
where $\alpha_{\rm ce}$ is an efficiency parameter and $\lambda$ a
parameter describing the strucure of the envelope of the giant. Both
are uncertain so we use them combined: $\alpha_{\rm ce} \; \lambda =
2$.

\subsubsection{Envelope ejection}\label{a:dynamic}

In the case of envelope ejection \citep{nvy+00}, we assume that the
complete envelope is lost and that this mass loss reduces the
  angular momentum of the system linearly proportional to the mass
  loss, as first suggested for the general case of non-conservative
  mass transfer by \citet{pz67}
\begin{equation}
J_{\rm i} - J_{\rm m} = \gamma J_{\rm i} \frac{\Delta M}{M_{\rm tot}},
\end{equation}
where $J_{\rm i}$ is the angular momentum of the pre-mass transfer
binary and $M_{\rm tot}$ is the total mass of the binary. The
companion does not accrete al all \citep[see discussion in
Sect.~\ref{masstransfer} and][]{nvy+00}.  The change in orbital
separation is given by
\begin{equation}\label{eq:orbit_change}
\frac{a_{\rm f}}{a_{\rm i}} = \left(\frac{M_{\rm f} \; m_{\rm
f}}{M_{\rm i} \; m_{\rm i}}\right)^{-2} 
\left(\frac{M_{\rm f} + m_{\rm f}}{M_{\rm i} + m_{\rm i}}\right) 
\left(1 - \gamma \frac{M_{\rm i} - M_{\rm f}}{M_{\rm i} + m_{\rm
i}}\right)^2.
\end{equation}
In this work we use $\gamma = 1.75$.

\subsubsection{Double spiral-in}\label{a:double_spi}

If mass transfer is unstable when both stars are evolved (which can
only happen if the mass ratio is close to unity), we model the
evolution as a common envelope in which the two cores spiral-in.  The
energy needed to expel the complete envelope is computed analogously
to the case of a standard common envelope \citep[see also
Sect.~\ref{a:stace}]{web84}:
\begin{displaymath}
\frac{M_{\rm i} \; (M_{i} - M_{\rm f})}{\lambda \; R} + 
\frac{m_{\rm i} \; (m_{\rm i} - m_{\rm f})}{\lambda r}
 = \alpha_{\rm ce} \; \left[
          \frac{M_{\rm f}\; m_{\rm f}}{2 \;a_{\rm f}} - 
          \frac{M_{\rm i}\; m_{\rm i}}{2 \;a_{\rm i}} \right]
\end{displaymath}
If the final separation is too small for the two cores to form a
detached binary, the cores merge and we compute the fraction of the
envelopes that is lost with the (practical) assumption that both
stars lose the same fraction of mass, retaining $f M$, i.e
\begin{displaymath}
\frac{M_{\rm i} (1-f) M_{\rm i}}{\lambda \; R} + 
\frac{m_{\rm i} (1-f) m_{\rm i}}{\lambda r}
 = \alpha_{\rm ce} \left[
          \frac{f M_{\rm i}\; f m_{\rm i}}{2 \;a_{\rm RLOF}} - 
          \frac{M_{\rm i}\; m_{\rm i}}{2 \;a_{\rm i}} \right]
\end{displaymath}
where $a_{\rm RLOF}$ is the  separation at which one of the 
cores  fills its Roche lobe. This is solved for $f$.

\end{document}